\documentclass[twocolumn,aps,showpacs,longbibliography,superscriptaddress]{revtex4}  
 \usepackage{graphicx}  
 \usepackage{graphics}
 \usepackage{dcolumn}   
 \usepackage{bm}        
 \usepackage{amssymb}   
 \usepackage{color}
 \usepackage{natbib}

\begin{document}

\title{Transcriptional bursting in gene expression: analytical results for general stochastic models}
\author{Niraj Kumar}
\affiliation{Department of Physics, University of Massachusetts Boston, Boston MA 02125, USA}
\author{Abhyudai Singh}
\affiliation{Department of Electrical and computer Engineering, University of Delaware, Newark, Delaware, USA}
\author{Rahul V. Kulkarni}
\affiliation{Department of Physics, University of Massachusetts Boston, Boston MA 02125, USA}
\date{\today}

\begin{abstract}
Gene expression in individual cells is highly variable and 
sporadic, often resulting in the synthesis of mRNAs and proteins
in bursts. Bursting in gene expression is known to impact 
cell-fate in diverse systems ranging from latency in HIV-1 
viral infections to cellular differentiation. It is generally 
assumed that bursts are geometrically distributed and that they 
arrive according to a Poisson process. On the other hand,
recent single-cell experiments provide evidence for complex burst
arrival processes, highlighting the need for more general stochastic
models. To address this issue, we invoke a mapping between general
models of gene expression and systems studied in queueing theory to
derive exact analytical expressions for the moments associated with
mRNA/protein steady-state distributions. These moments are then used
to derive explicit conditions, based entirely on experimentally
measurable quantities, that determine if the burst distributions
deviate from the geometric distribution or if burst arrival deviates
from a Poisson process.  For non-Poisson arrivals, we develop
approaches for accurate estimation of burst parameters.
\end{abstract}
\maketitle

\section{Introduction}

The cellular response to fluctuating environments requires adjustments to cellular phenotypes driven by underlying changes in gene expression. Given the inherent stochasticity of cellular reactions, biological circuits controlling gene 
expression have to operate in the presence of significant noise \cite{elowitz2002stochastic,kaern2005stochasticity,raser2005noise,sanchez2013regulation,eldar2010functional,
raj2008nature,larson2011expression,junker2014every,munsky2012using,golding2011decision,bar2006noise,newman2006single,weinberger2012expression,kumar2014exact}. 
While noise reduction and filtering is
essential for several cellular processes \cite{PhysRevX.4.041017}, cells can also amplify and utilize intrinsic noise to generate phenotypic 
diversity that enables survival under stressful conditions \cite{Balazsi2011Cell}. Recent studies have demonstrated the 
importance of such bet-hedging survival strategies in diverse processes ranging from viral infections to bacterial competence \cite{Balazsi2011Cell}.
Quantifying the kinetic mechanisms of gene expression that drive variations in a population of cells will thus contribute towards a fundamental understanding of cellular functions with important applications to human health.

Recent experiments focusing on gene expression at the single-cell level have revealed
striking differences from the corresponding population-averaged
behavior. In particular, it has been demonstrated that
transcription in single cells is sporadic, with mRNA synthesis often
occurring in bursts followed by variable periods of inactivity\cite{suter2011origins,coulon2013eukaryotic,Golding20051025,chubb2006transcriptional,raj2006stochastic,so2011general,
larson2011expression,taniguchi2010quantifying,zong2010lysogen,sanchez2013genetic,dar2012transcriptional,singh2012dynamics}. Such
transcriptional bursting can give rise to high variability in gene
expression products and to phenotypic variations in a population of
genetically identical cells \cite{gefen2008single,weinberger2005stochastic,Zeng2010682,wernet2006stochastic}. Furthermore, dynamical parameters that characterize
transcriptional bursting of key genes can significantly influence
cell-fate decisions in diverse processes ranging from HIV-1 viral infections to stem-cell differentiation \cite{Balazsi2011Cell}. Correspondingly, there is significant interest
in developing approaches for quantifying parameters related to
transcriptional bursting such as frequency and mean burst size.

In recent years, multiple studies have provided evidence for bursty synthesis of
mRNAs \cite{Golding20051025,chubb2006transcriptional,raj2006stochastic,so2011general,
taniguchi2010quantifying,zong2010lysogen,ochiai2014stochastic,senecal2014transcription} and proteins \cite{cai06,Xie2006}. 
Experimental approaches in such studies include both
steady-state measurements and time-dependent measurements of the mean
and variance of gene expression products at the single-cell level.
While obtaining time-lapse measurements of bursts at the single-cell level can be challenging, steady-state measurements at the single-cell level are now
carried out routinely.  It would thus be desirable to
develop approaches for making inferences about burst parameters in
gene expression using steady-state measurements at the single-cell
level.

As noted in \cite{Pedraza2008}, steady-state measurements
of the mean and variance alone cannot be used for estimating burst
parameters for general models of gene expression, e.g. when burst arrival 
is governed by complex promoter-based regulation \cite{zhang2014promoter}.  
Additional insights into processes leading to transcriptional 
bursting can potentially be obtained using  measurements of higher moments. 
However, analytical results for higher moments of
steady-state mRNA and protein distributions in general models of expression 
have not been obtained so far. The derivation of the corresponding analytical expressions will elucidate how measurement of higher moments 
can potentially lead to quantification of burst parameters. 
To address these issues, it is essential to develop and
analyze a general class of stochastic models of gene expression.

A simple stochastic model that is widely used in analyzing bursting in gene
expression is the random telegraph model that takes into account the
switching of promoter between transcriptionally active (ON)
and inactive (OFF) states \cite{peccoud1995markovian,shahrezaei2008analytical,dobrzynski2009elongation}. This model has been used as the
basis for several studies focusing on inferring gene expression
parameters based on observations
of the mean and variance of mRNA/protein distributions \cite{skupsky2010hiv,dar2012transcriptional,weinberger2012expression}.
In this model, in the limit that we have transcriptional bursting, the
arrival of bursts is a Poisson process. Correspondingly, the    
waiting-time distribution between arrival of mRNA bursts is assumed to
be exponential. In general, this assumption is not valid as there are
multiple kinetic steps involved in promoter activation \cite{PhysRevLett.106.058102,Pedraza2008,xu2013stochastic}. 
Recent experiments on mammalian genes \cite{suter2011mammalian,harper2011dynamic,larson2011expression} have demonstrated
that the waiting-time for arrival of bursts
does not have an exponential distribution. In view of these
experimental observations, it is natural to ask: Using steady-state measurements can we infer if 
the burst arrival process is {\em not} a Poisson process? If so, how can we estimate the corresponding burst parameters?

Furthermore, in estimating burst size it is commonly
 assumed that mRNA/protein bursts are geometrically distributed. This
 assumption, which has been validated by experimental observations for some 
genes,  is derived from the corresponding distribution of bursts in the random
telegraph model.  
However, given the complexity and diversity of gene expression mechanisms,
it is possible that several promoters involve multiple rate-limiting steps 
in the transition from the ON state to the OFF state. 
In such cases, the transcriptional burst size distribution will not be 
a geometric distribution.  This observation leads to the following
question: Can we use steady-state measurements of moments to determine
if the burst distribution deviates from a geometric distribution?

The aim of this paper is to address the above questions
  by considering models with general arrival processes for mRNA
  creation. The paper is organized as follows. In
  Sec. \ref{sec:model}, we introduce a class of gene expression models
  with general arrival processes leading to mRNA/protein bursts with
  arbitrary burst distribution. In Sec. \ref{sec:mapping}, we review
  the mapping from gene expression models to systems studied in
  queuing theory \cite{PhysRevLett.106.058102,KulkarniPRE2010,liu00} and use it to derive
  steady-state moments for mRNA/protein distributions.  In
  Sec \ref{sec:geometric}, we use the analytical expressions obtained for
  the steady-state moments to derive a condition for determining if
  the distribution of mRNA bursts is geometric and illustrate the
  condition derived using exactly solvable models. In
  Sec. \ref{sec:poisson}, we derive conditions that determine if the
  arrival of mRNA bursts deviates from a Poisson process and
  illustrate these conditions derived using simple models. For models
  with non-Poissonian arrival of mRNA bursts, a method for estimating
  burst size is presented in sec. \ref{sec:burst}. Finally,
  Conclusions appear in Sec. \ref{sec:conclusions}.  
  

\section{Model and Preliminaries}
\label{sec:model}

\begin{figure}
\centering
\includegraphics[width=7cm]{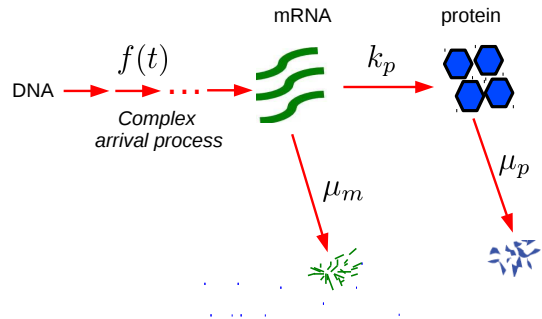}
\caption{Kinetic scheme for the gene expression with general arrival time distributions, $f(t)$, for  mRNA bursts, which in 
turn can produce proteins with rate $k_p$. Both mRNAs and proteins decay with rates $\mu_m$ and $\mu_p$, respectively.}
\label{fig:model}
\end{figure}

We consider a general model of gene
expression as outlined in Fig.\ref{fig:model}. In the model, mRNAs are
produced in bursts, with $f(t)$ representing a general arrival time
distribution for mRNA bursts. The mRNA burst distribution can be arbitrary. 
Each mRNA then produces proteins with rate $k_p$, and
finally, both mRNAs and proteins decay with rates $\mu_m$ and
$\mu_p$, respectively.  Note that the
model also allows for post-transcriptional regulation since the protein
burst distribution from each mRNA can be arbitrary; the only
assumption is that each mRNA produces proteins independently.

In the limit $\mu_p \ll \mu_m$, we can use the bursty synthesis approximation
\cite{shahrezaei2008analytical} for analyzing protein dynamics. This approximation consists of two steps: 
1) obtaining
the distribution of proteins produced from each mRNA and 2) assuming
that the proteins are produced in instantaneous bursts. The corresponding 
distribution for the number of proteins created is referred to as the protein burst 
distribution. A detailed justification of the validity of this approximation has been provided in 
previous work \cite{shahrezaei2008analytical,bokes2012multiscale}.

Let $a^p(z)=\sum_{n=0}^{\infty}z^np(n)$ denote the generating function of the protein burst
distribution $p(n)$ produced by a {\em single} mRNA, and let $A^p(z)=\sum_{n=0}^{\infty}z^nP(n)$ denote
the generating function of the protein burst distribution $P(n)$ produced by {\em all} the
mRNAs in a burst. If we denote by $A^m(z)$ the generating function of
the mRNA burst distribution, then we have the following relation
between the generating functions
\begin{equation}{\label{gen_fn_eqn}}
A^p(z) = A^m\left[a^p(z)\right].
\end{equation}
The above relation follows from the observation that the number of
proteins produced in a burst is a compound random variable: the sum
of $m$ independent identical random variables, each of which
corresponds to the number of proteins produced from a single mRNA in
the burst and $m$ itself is a random variable denoting the number of
mRNAs produced in the burst.

While the analytical results that we derive are valid for general mRNA
and protein burst distributions, we will primarily focus on a
specific class of burst distributions. Simple kinetic models and the
results from multiple experiments indicate that mRNA burst
distributions are geometric \cite{cai06}. Similarly, the burst distribution of
proteins produced from a single mRNA is a geometric distribution with
mean $\langle p_b \rangle=k_p/\mu_m$. For a geometric distribution with mean
$\langle p_b \rangle$, the
generating function is given by  
$$a^p(z) = \frac{1}{\left[1 + \langle p_b \rangle(1-z)\right]}.$$

If we condition the geometric distribution on the production of at least 1
mRNA, then the generating function for the corresponding conditional
geometric distribution is given by $$A^m(z) = \frac{z}{\left[1+ \langle m_b \rangle(1-z)\right]}$$ 
with $(1 + \langle m_b \rangle)$ as the mean mRNA burst size.  Note that
in the limit $\langle m_b \rangle\to 0$, this distribution reduces to exactly $1$ mRNA
produced in each burst. Thus the conditional geometric distribution provides a unified representation 
of both Poisson arrival process for mRNAs ($\langle m_b \rangle\to 0$) and processes leading to 
transcriptional bursting ($\langle m_b \rangle > 0$).

Consider now the protein burst distribution produced by an underlying
conditional geometric mRNA burst distribution with mean $ (1 + \langle m_b \rangle)$. Using Eq. (\ref{gen_fn_eqn}), we see that the corresponding generating function of the 
protein burst distribution is given by 
$$ A^p(z)=\frac{1}{1+\langle m_b \rangle\langle p_b \rangle(1-z)}.$$
This is the generating function for a geometric distribution with mean
$b = (1+ \langle m_b \rangle)\langle p_b \rangle$, 
where $\langle p_b \rangle=k_p/\mu_p$ represents the mean protein 
burst size from a single mRNA.

Single-cell experiments have demonstrated that the protein burst mean $b$ can be 
directly measured in some cases \cite{cai06}. However, if the protein production rate $k_p$ is
not known, the preceding analysis implies that measurements of protein burst
distributions (which determine $b$) cannot be used to determine the
degree of transcriptional bursting $(1+\langle m_b \rangle)$. Since the mean 
transcriptional burst size is an important parameter characterizing bursting,
it is of interest to develop approaches for estimating it based on available 
experiments. Previous work \cite{ingram08} has
argued that the mean transcriptional burst size cannot be determined
using measurements of protein burst distributions alone or by using only protein
steady-state distributions. It was suggested that combining such
measurements can potentially provide a way of determining the mean
transcriptional burst size. To explore this possibility, it is
necessary to derive analytical results connecting moments of burst and steady-state
distributions for general kinetic schemes.

\section{Mapping to queueing theory: Results for moments}
\label{sec:mapping}
To obtain steady-state moments for the model outlined in Fig. 1, we invoke the mapping 
of this gene expression model to systems studied in queueing theory \cite{PhysRevLett.106.058102,cookson2011queueing,mather2010correlation,liu00}. Broadly speaking, queueing theory is the mathematical theory of waiting lines formed by customers who, arriving according some random protocol, stay in the system until they receive service from a group of servers. 
Such queues are typically characterized by specifying a) the stochastic process 
governing the  arrival of customers, b) distribution of number of customers in each arrival, c) the stochastic process governing departure of customers, and 
d) the number of servers. 
When the gene expression model in Fig. 1 is expressed in the language of queueing theory, individual mRNAs/proteins are the 
analogs of customers in queueing models. The production of mRNAs/proteins in bursts corresponds to the arrival of
customers in batches. Just as the customers leave the queue after receiving service, mRNAs/proteins exit the system upon degradation. Thus the 
waiting-time distribution for mRNA/protein decay is the analog of 
service time distribution for customers in queueing models.
For the model in Fig. 1, their decay time distribution is the exponential
distribution. Also, since  mRNAs/proteins are degraded independently of each other, the corresponding number of servers in queueing models is $\infty$ (which ensures that presence of a customer in the system does not affect the 
service time of any other customer in the system).

Based on above mapping, the queueing system corresponding to the model outlined in Fig. 1 is the $GI^X/M/\infty$ system \cite{PhysRevLett.106.058102,liu00}. In this model, the symbol $G$ refers to a general
waiting-time distribution for the arrival process, $I^X$ denotes
customers arriving in batches of independently distributed random
sizes $X$, $M$ stands for Markovian (i.e. exponential) service-time distribution for
customers and $'\infty'$ stands for infinite servers.

For the $GI^X/M/\infty$ model, exact results for iteratively obtaining the 
moments of the steady-state distribution of the number of customers 
have been derived \cite{liu00}. Using these results, explicit expressions for the first four moments of the steady-state distribution are provided in the Appendix. Applying the mapping discussed above, these results can be translated into 
exact expressions for the moments of mRNA/protein steady-state distributions, 
as discussed below.

Let us first examine the expressions for steady-state means of mRNAs, $\langle m_s \rangle$, and proteins, $\langle p_s \rangle$, 
which are given by
\begin{equation}{\label{emeanMP}}
 \langle m_s\rangle=\frac{k_b}{\mu_m}\langle m_b \rangle, \hspace{0.2cm} \langle p_s \rangle=\frac{k_b}{\mu_p}b,
\end{equation}
where $k_b$ stands for the mean arrival rate of mRNA bursts and $b=\langle m_b \rangle  \langle p_b \rangle$ is 
the mean of the protein burst distribution (including contributions from all the mRNAs). 
Although Eq. (\ref{emeanMP}) has been derived by assuming that the arrival of mRNAs/proteins is a renewal 
process, it is valid for arbitrary arrival processes.
This is because Eq. (\ref{emeanMP}) is a direct consequence of Little's Law \cite{little1961proof,KulkarniPRE2010} which is valid for general arrival processes. 

The above equations, Eq. (\ref{emeanMP}), can be used to determine the mean transcriptional burst size, provided the protein burst distribution can be measured experimentally.
To see this, we note that dividing the expressions for the mean mRNA and protein levels leads to
\begin{equation}
\frac{b}{\langle m_b \rangle} = \frac{\mu_p}{\mu_m} \frac{\langle p_s \rangle} {\langle m_s \rangle}.
\end{equation}
Since the steady-state means $\langle m_s \rangle$ and $\langle p_s \rangle$ as well as the
degradation rates $\mu_m$ and $\mu_p$ are parameters that can be
measured experimentally, the above equation implies that
the ratio $b/\langle m_b \rangle$ can be determined experimentally.
Given $ b/\langle m_b \rangle= k_p/\mu_m$, this implies that the mean protein production rate $k_p$ can also
be determined experimentally. This is an important result since it provides an approach for determining  the 
mean protein production rate $k_p$  that is valid for arbitrary 
arrival processes for mRNAs. Furthermore, the above equation implies that, if the mean of protein burst distribution $b$ can 
be measured \cite{Xie2006}, then the mean transcriptional burst size $\langle m_b\rangle$ can also be determined.
Thus, if we have measurements 
for mean mRNA and protein numbers and also the mean of protein burst distribution,
then these measurements can be used to determine the degree of transcriptional bursting $\langle m_b \rangle$
as well as the parameters $\langle p_b \rangle$ and $k_p$. It is noteworthy that this procedure for 
estimating the burst parameters is valid for arbitrary stochastic processes corresponding to mRNA transcription.

We next turn to expressions for higher moments of mRNA and protein steady-state distributions.
The noise in mRNA steady-state distributions is given by
\begin{eqnarray}{\label{enoiseM}}
 \frac{\sigma^2_{m_s}}{\langle m_s\rangle ^2}&=&\frac{1}{\langle m_s\rangle}+\frac{\mu_m}{k_b}+\frac{\mu_m}{2k_b}\left[K_g(\mu_m)-1
+\frac{\sigma^2_{m_b}}{\langle m_b \rangle ^2}\right.\nonumber\\&&-\left.\left(1+\frac{1}{\langle m_b \rangle}\right)\right],
\end{eqnarray}
where $\sigma^2_{m_b}$ is the variance of mRNA burst distribution and $K_g(\mu_m)$ is the so-called gestation factor, 
\begin{equation}{\label{ekg}}
K_g(\mu_m)=1+2\left[\frac{f_L(\mu_m)}{1-f_L(\mu_m)}-\frac{k_b}{\mu_m}\right],
\end{equation}
with $f_L(s)$ denoting the Laplace transform of arrival time distribution of mRNA bursts. The function $K_g(\mu_m)$ encodes information about the arrival process. Specifically, we note that for Poisson arrivals, 
we have $K_g(\mu_m)=1$. 

For proteins (in the burst limit $\mu_m\gg\mu_p$), we obtain \cite{PhysRevLett.106.058102}
\begin{eqnarray}{\label{enoiseP}}    
 \frac{\sigma^2_{p_s}}{\langle p_s\rangle ^2}&=&\frac{1}{\langle p_s \rangle}+ \frac{\mu_p}{k_b}+ \frac{\mu_p}{2k_b}\left[K_g(\mu_p)-1+
 \frac{\sigma^2_{m_b}}{\langle m_b \rangle ^2}\right.\nonumber \\ &&\hspace{-1.cm}-\left.\left(1-\frac{1}{\langle m_b \rangle}\right)
+\left( \frac{\sigma_{p_b}^2}{\langle p_b\rangle^2}
 -\left(1+\frac{1}{\langle p_b\rangle}\right)\right)\frac{1}{\langle m_b \rangle}\right],\nonumber\\
 \end{eqnarray}
where $K_g(\mu_p)$ is given by Eq.(\ref{ekg}) and $\sigma^2_{p_b}$ is the variance of protein burst distribution produced by a single mRNA. 
The expression for protein noise is composed of the noise term for the
basic two-stage model of gene expression \cite{shahrezaei2008analytical} and additive noise
contributions due to: a) deviations from exponential waiting-time
distribution for the arrival process, b) deviations from conditional geometric
distributions for mRNA burst distributions and c) deviations from geometric
distributions for protein burst distributions.
For both mRNAs and proteins, the noise in steady-state distributions depends on all the moments of the burst arrival time distribution through the term $K_g$. Therefore, arrival processes corresponding  to different kinetic schemes for transcription will make different contributions to the overall noise,
even if they have identical means and variances for the the burst arrival time distribution.

We note from Eq. (\ref{enoiseM}) that, for Poisson arrivals, i.e. $K_g=1$, and geometrically distributed burst, i.e. $\sigma^2_{m_b}=\langle m_b\rangle(\langle m_b\rangle- 1$), the equations for the noise and mean have only two unknown burst parameters, $k_b$ and $\langle m_b\rangle$. In this case, experimental measurements of the first two moments of the steady-state distribution are sufficient to estimate the burst parameters, as has been done in multiple studies.
However, when the arrival process is non-Poisson or if the burst distribution deviates from a geometric distribution, measurements of the first two steady-state moments are not sufficient for estimating the burst parameters. This observation 
motivates the need for analytical expressions for the higher moments which we turn to next.

We now derive analytical expressions for the third moment, specifically the skewness parameter.
For mRNAs, the exact expression for skewness $\gamma_{m_s}$ is given by
\begin{eqnarray}{\label{eskewM}}
 \frac{\gamma_{m_s}\sigma^3_{m_s}}{m_s}&=&1+\langle m_s \rangle \langle m_b \rangle\mathcal{K}_1(\mu_m)
+2\langle m_b\rangle^2\mathcal{K}_2(\mu_m,\langle m_b \rangle)\nonumber\\&+&\left(\sigma^2_{m_b}+\langle m_b\rangle^2-\langle m_b \rangle\right)\mathcal{K}_3(
\mu_m,\langle m_b \rangle)\nonumber\\&+&\frac{\langle m_b(m_b-1)(m_b-2)\rangle}{3\langle m_b\rangle},
\end{eqnarray}
where we have defined
\begin{eqnarray}{\label{ekappa}}
 \mathcal{K}_1(\mu_m)&=&K_g(2\mu_m)-K_g(\mu_m),\nonumber\\ 
 \mathcal{K}_2(\mu_m,\langle m_b\rangle)&=&\frac{K_g(\mu_m)-1}{4}\left(\frac{3}{\langle m_b\rangle}+K_g(2\mu_m)-1\right),\nonumber\\
 \mathcal{K}_3(\mu_m,\langle m_b\rangle)&=&\frac{3}{2\langle m_b\rangle}+\frac{K_g(\mu_m)+K_g(2\mu_m)}{2}-1.
 \end{eqnarray}
For proteins, we obtain in the burst-limit ($\mu_m\gg\mu_p)$,
\begin{eqnarray}{\label{eskew1P}}
  \frac{\gamma_{p_s}\sigma^3_{p_s}}{p_s}&=&1+(A_1^p)^2\left[\frac{\langle p_s \rangle}{b_{}}\mathcal{K}_1(\mu_p)+2\mathcal{K}_2(\mu_p,A_1^p)
\right]\nonumber\\&+&A_2^p\mathcal{K}_3(\mu_p,A_1^p)+\frac{A_3^p}{3A_1^p},
\end{eqnarray}
where $\mathcal{K}_1, \mathcal{K}_2, \mathcal{K}_2$ are given using Eq. (\ref{ekappa}), $A_k^p$ is given by $A_k^p=d^kA^p(z)/dz^k|_{z=1}$ 
and using Eq. (\ref{gen_fn_eqn}) \cite{Ross:2006:IPM:1197141} 
we obtain the parameters $A_k^p$ as:
\begin{eqnarray}{\label{ekp}}
 A_{1}^p&=&\langle m_b \rangle \langle p_b \rangle,\nonumber\\
A_{2}^p&=&\langle m_b \rangle \left(\sigma^2_{p_b}-\langle p_b \rangle\right)+\left( \sigma^2_{m_b}+\langle m_b \rangle ^2\right)\langle p_b\rangle^2,\nonumber\\
A_{3}^p&=&\langle p_b\rangle^3\langle m_b(m_b-1)(m_b-2)\rangle+3\langle m_b(m_b-1)\rangle\langle p_b\rangle\nonumber\\&&\langle p_b(p_b-1)\rangle+\langle m_b \rangle
 \langle p_b(p_b-1)(p_b-2)\rangle.
\end{eqnarray}
Similarly, expressions for higher order moments of protein and mRNA steady-state distributions can be obtained iteratively. 
The corresponding expressions for the kurtosis are provided in the Appendix A.

The analytical results derived above for proteins are exact in the burst limit, which assumes that proteins are produced instantaneously from all the mRNAs in a burst. Going beyond the burst limit (i.e. not limited to $\mu_m\gg\mu_p)$, exact results for the higher moments of the protein steady-state distribution will, in general, depend on the details 
of the kinetic scheme for gene expression. However, we can derive approximate analytical expressions for general schemes by requiring that: 
a) the results reduce to the exact results in the burst limit and b) they match the exact results for the two-stage model of gene expression.

For the two-stage model, exact results for the first four moments have been derived by Bokes et. al\cite{bokes2012exact}. Comparing these exact results with our results derived in 
the burst limit, we observe that results of \cite{bokes2012exact} can be reproduced by a suitable scaling of the burst-size parameters $A_k^p$. 
For example, the exact expression for the noise is obtained by the following scaling \cite{PhysRevLett.106.058102}.

\begin{equation}{\label{enoisepCorr}}
\left(\frac{\sigma^2_{p_s}}{\langle p_s \rangle^2}-\frac{1}{\langle p_s \rangle}\right)\rightarrow \left(\frac{\sigma^2_{p_s}}{\langle p_s \rangle^2}
-\frac{1}{\langle p_s \rangle}\right)\frac{1}{1+\frac{\mu_p}{\mu_m}}.
\end{equation}
Similarly, for the expression for skewness, the parameters $A_2^p$ and $A_3^p$ are scaled as: 
\begin{equation}{\label{eA2A3}}
 A_2^p\rightarrow A_2^p\frac{1}{1+\frac{\mu_p}{\mu_m}}\hspace{0.2cm} \text{and} \hspace{0.2cm} A_3^p\rightarrow A_3^p\frac{1}{(1+\frac{\mu_p}
{\mu_m})(1+2\frac{\mu_p}{\mu_m})}.
\end{equation}
As shown in Fig.2b the resulting analytical expressions using this approach show good agreement with results from simulations \cite{gillespie1977exact}.
\begin{figure}[h]
 \centering 
\includegraphics[width=9cm]{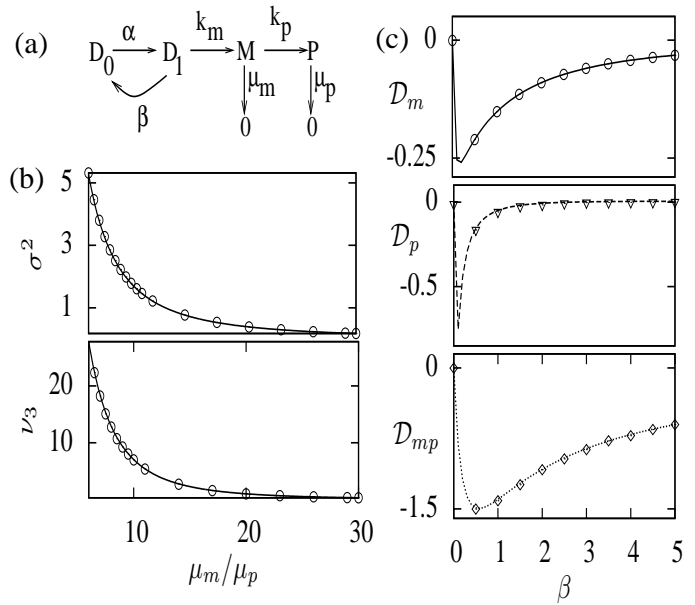}
 \caption{Kinetic scheme for the two-state random telegraph model has been shown in (a).
 Here promoter for the gene switches from OFF ($\text{D}_0$) to ON ($\text{D}_1 $) state with rate $\alpha$ and that from ON to OFF with rate $\beta$. The promoter in the ON state
gives rise to conditional geometric mRNA bursts of mean size $\langle m_b \rangle$ with rate $k_m$, which in turn can  produce proteins with rate $k_p$. Both mRNAs and proteins can degrade with rates 
$\mu_m$ and $\mu_p$, respectively. For this model, 
 steady state variance (scaled by $10^{-5}$) and third central moment $\nu_3$ (scaled by $10^{-6}$) 
of proteins 
are plotted as a function of $\mu_m/\mu_p$
in (b): lines represent analytic estimates while points correspond to the simulation results. Parameters are: $\alpha=0.5$, $\beta=0.25$, $k_m=2$, $\langle m_b \rangle=5$, $k_p=0.5$. 
In (c), signatures for non-Poisson arrival $\mathcal{D}_m$, $\mathcal{D}_m$ and $\mathcal{D}_m$  are plotted for the model shown in (a) as a function 
of {\emph off} rate $\beta$: analytic estimates are shown as lines while points correspond to the simulation results with 
parameters: $\alpha=0.25$, $k_m=2$, $\langle m_b \rangle=5$, $k_p=0.5$, $\mu_m=1$, $\mu_p=0.01$.}
\label{fig:23moments}
\end{figure}

It is noteworthy that the results derived are valid for a general
class of kinetic schemes of gene expression. For a specific kinetic
scheme, we can determine the corresponding waiting-time distribution
for the arrival process and the burst distributions for mRNA and
proteins. Substituting these results in the equations derived leads to
the corresponding expressions for the moments of the steady-state
distribution. The results obtained can thus provide insight into how
specific kinetic schemes of gene expression (e.g. combining
promoter-based regulation and post-transcriptional regulation) can be
used to impact the noise and higher moments of steady-state
distributions.

\section{Signatures for non-geometric bursts}
\label{sec:geometric}
The expressions derived for the mRNA/protein steady-state moments indicate that the estimation of burst 
parameters depends on accurate representation of the burst size distributions and the burst arrival 
time distribution.  It is widely assumed that the mRNA burst distribution can be represented by a 
conditional geometric distribution (i.e. including both single mRNA arrivals and geometrically 
distributed burst arrivals). While this assumption is consistent with multiple experimental 
observations, for general kinetic schemes the possibility of non-geometric mRNA burst distributions 
has to be considered.

To address the possibility of non-geometric mRNA burst distributions, let us first consider that the random variable corresponding to the mRNA burst distribution ($m_b$) 
has a conditional geometric distribution. That is, the probability that a burst produces $n$ mRNA molecules is given by
\begin{equation}{\label{econdg}}
 {\mathcal {P}}(m_b=n)=(1-p)^{n-1}p,
\end{equation}
where $0<p\le1$, and $n=1,2,3\dots\infty$.
This distribution leads to
\begin{equation}{\label{evarmb}}
\sigma_{mb}^2=\langle m_b\rangle(\langle m_b \rangle-1).
\end{equation}
Using Eqs.~(\ref{emeanMP}) and (\ref{evarmb}) in Eq.~(\ref{enoiseM}), and denoting $F_m=\sigma_{m_s}^2/\langle m_s \rangle$ as the Fano 
factor of mRNA copy numbers, Eq.~(\ref{enoiseM}) can be rewritten as
\begin{equation}{\label{efanoM}}
 F_m(\mu_m)=\frac{\langle m_b \rangle}{2} \left[1+K_g(\mu_m)\right].
\end{equation}
Similarly, using the burst size distribution from Eq.~(\ref{econdg}), the skewness in Eq.~({\ref{eskewM}) is given by
\begin{eqnarray}{\label{eskewM1}}
 \frac{\gamma_{m_s}\sigma^3_{m_s}}{\langle m_s \rangle}&=&1+\langle m_s \rangle \langle m_b \rangle\mathcal{K}_1(\mu_m)
       +2\langle m_b \rangle^2\mathcal{K}_2(\mu_m,\langle m_b \rangle)\nonumber\\
       &+&2\left(\langle m_b \rangle-1\right)\left[\left(1+\mathcal{K}_3(\mu_m,\langle m_b \rangle)\right)\langle m_b \rangle
       -1\right].\nonumber\\
\end{eqnarray}

We note that Eq.(\ref{eskewM1}) connects experimentally measurable moments of the steady-state distribution to the parameters $K_g(\mu_m)$,
$K_g(2\mu_m)$ and $\langle m_b \rangle$. 
Furthermore, note that Eq. (\ref{efanoM}) can be recast as $K_g(\mu_m)=(2F_m(\mu_m)/\langle m_b \rangle)-1$.
Now, considering a change in the degradation rate from $\mu_m$ to $2 \mu_m$ (keeping the mean
burst size, $\langle m_b \rangle$ invariant), we obtain
\begin{equation}
K_g(2\mu_m)=(2F_m(2\mu_m)/\langle m_b \rangle)-1.  
\end{equation}
Using the above in Eq.~(\ref{eskewM1}),
we get an expression connecting experimentally measurable quantities
associated with moments of the mRNA steady-state distribution. 
The resulting expression is:
\begin{eqnarray}{\label{eGm}}
 \mathcal{G}_m&\equiv&\frac{\gamma_{m_s}\sigma^3_{m_s}\langle m_s \rangle^{-1}}{2F_m(2\mu_m)(\langle m_s \rangle-1)
  +F_m\left(1-2\langle m_s \rangle+2F_m(2\mu_m)\right)}\nonumber\\&=&1.
\end{eqnarray}

We note that the above expression has been derived by making just one
assumption, namely, the mRNA burst distribution is a conditional geometric distribution. The derived expression thus indicates that a combination of experimentally measurable quantities has to deviate from 1 if the mRNA burst distribution deviates from a conditional geometric distribution. Thus the analytical results derived provide a signature for deviation from conditional geometric mRNA bursts using measurements of the first three moments of the mRNA steady-state distribution.

The main requirement for using the above relation is that measurements of mRNA steady-state distribution can be carried out at two different rates of the mRNAs $\mu_m$ and $2\mu_m$. 
Given that mRNA degradation rates can be tuned experimentally, a   
straightforward strategy to ensure that the degradation rate is tuned to
twice the original value ($2\mu_m$) is to compare the mean mRNA levels at $\mu_m$ and $2\mu_m$.
Given these measurements, a value of
$\mathcal{G}_m\ne1$ implies that bursts are not distributed
geometrically. The strength of this result lies in the fact that it
holds for general arrival processes for mRNA bursts with arbitrary waiting-time distributions.

Let us consider a specific simple model to illustrate the condition derived above.
First, let the arrival process for mRNA bursts be a Poisson process. For this, arrival time 
distributions of mRNA bursts in the time domain, $t$, and in the Laplace domain, $s$, are given by
\begin{equation}{\label{n1}}
 f(t)=k_be^{-k_bt},~~~~~f_L(s)=k_b/(k_b+s),
\end{equation}
where $k_b$ is the rate of arrival of mRNA bursts. For the mRNA burst distribution, let us assume that it is given by the negative binomial distribution, i.e.
\begin{equation}{\label{n2}}
 \mathcal{P}(m_b=n)=\frac{(n + r - 1)!}{n!(r-1)!}p^n(1-p)^r,
\end{equation}
where $0<p\le1$, $r\ge 1$, and $n=0,1,2,3\dots\infty$.  For $r=1$, the above reduces to the geometric distribution 
and therefore we expect  $G_m=1$ in this limit. Using the expressions for the moments derived in Sec \ref{sec:mapping},
we obtain an explicit expression for  $G_m$ (Appendix B):
\begin{equation}
 \mathcal{G}_m=\frac{1}{3} \left(-\frac{p+1}{p r+1}+\frac{4}{p (r-1)+2}+2\right).
\end{equation}
Notice that for the geometric bursts ($r=1$) we get $\mathcal{G}_m=1$, as expected. However, for non-geometric bursts, 
deviations of $\mathcal{G}_m$ from 1 are observed (also see Fig.\ref{fig:nb} in Appendix). 
Two additional examples of microscopic models for non-geometric bursts (the two state random telegraph model and
a model with three promoter states where mRNAs are produced from two states) are discussed in the Appendix.

The preceding analysis can be extended to protein steady-state distributions to 
derive a similar condition for geometric burst distributions in terms of
steady state moments associated with proteins (see Appendix). 

\section{Signatures for non-Poisson arrivals}
\label{sec:poisson}

The analytical expressions derived for the steady-state moments for mRNAs and proteins can also be used to 
make inferences about the burst arrival process based on steady-state measurements. 
Since multiple studies assume that the burst arrival process is characterized by an exponential waiting-time 
distribution, it would be useful to determine if this assumption is invalid using measurements of steady-state 
distributions. As shown below, we can obtain conditions for the same using the results derived for 
higher moments.

In the following, we will focus on the cases that the mRNA burst distribution is conditional geometric
and the protein burst distribution is geometric, which is consistent with multiple experimental observations. As discussed, 
choosing the conditional geometric distribution for mRNAs allows us to consider both single mRNA arrivals 
and geometric mRNA bursts in one framework.
Since experiments can provide measurements of both mRNA and protein steady-state distributions, it is useful to have conditions 
for the arrival process using either mRNA data or protein data or both mRNA and protein data.
Based on these three possibilities, we present three different conditions in the following.

\subsection{Using moments of mRNA steady-state distributions}
Let us first consider the case where we have only measurements of the mRNA steady-state distribution. We note that for Poisson arrivals $K_g(\mu_m)=K_g(2\mu_m)=1$,
and using the expressions for mean and noise from Eqs. (\ref{emeanMP}) and (\ref{enoiseM}) we get, $F_m=\langle m_b \rangle$, where
 $F_m=\sigma^2_{m_s}/\langle m_s\rangle$ is the mRNA Fano factor. Further, using this in 
the equation for skewness, Eq.~(\ref{eskewM}), we derive the following condition that must be satisfied if  
the arrival of mRNA bursts is a Poisson process.
\begin{equation}{\label{econditionM}}
 \mathcal{D}_m\equiv\frac{\gamma_{m_s}\sigma^3_{m_s}}{\langle m_s \rangle\left[3(F_m-1)\left\{1+\frac{2}{3}(F_m-1)\right\}+1\right]}-1=0.
\end{equation}
Thus $\mathcal{D}_m\ne 0$ is a signature of non-Poisson arrival processes. Since the above prescription is based on experimentally measurable 
quantities such as $\langle m_s \rangle, \sigma^2_{m_s},\gamma_{m_s}$ and $\mu_m$, it can be used to determine if the assumption 
of a Poisson arrival process is invalid. 

\subsection{Using moments of protein steady-state distributions}
We next consider the case where we have access to only the protein steady-state distribution. The steps followed are 
similar to those outlined for the mRNA case. For Poisson arrivals, $K_g(\mu_p)=1$, and using Eq.~(\ref{emeanMP})and 
(\ref{enoisepCorr}) we get 
$$b=(F_p-1)\left(1+\frac{\mu_p}{\mu_m}\right),$$  where $F_p=\sigma^2_{p_s}/\langle p_s\rangle$ is the protein Fano factor. Substituting this 
in the expression for protein skewness, Eq.~(\ref{eskew1P}) with the scaled $A_2^p$ and $A_3^p$ given by Eq. (\ref{eA2A3}), we arrive at the following condition for Poisson arrivals. 
\begin{eqnarray}{\label{econditionP}}
      \mathcal{D}_p&\equiv&\frac{\gamma_{p_s}\sigma^3_{p_s}}{\langle p_s\rangle\left[3(F_p-1)\left\{1+\frac{2}{3}\left(\frac{\mu_m+\mu_p}{\mu_m+2\mu_p}
                    \right)(F_p-1)\right\}+1\right]}-1\nonumber\\&=&0
      \end{eqnarray} 
Again, non-zero value of $\mathcal{D}_p$ is a signature of non-Poisson arrivals. 

\subsection{Using both mRNA and protein steady-state distributions}
Finally, if we have both mRNA and protein steady-state distribution measurements available,
then the condition for Poisson arrivals can be obtained by combining measurements of second moments of mRNA and protein distributions as follows: 
Using Eqs.~(\ref{emeanMP}),(\ref{enoiseM}) and (\ref{enoisepCorr}), we get, 
\begin{eqnarray}{\label{econditionMP}}
 \mathcal{D}_{mp}&\equiv&\frac{F_m}{\langle m_s\rangle\mu_m}-\frac{(\mu_m+\mu_p)(F_p-1)}{\mu_m\langle p_s \rangle\mu_p}
 \nonumber\\&=&\frac{1}{2k_b}\left[K_g(\mu_m)-K_g(\mu_p)\right],
\end{eqnarray}
which vanishes for Poisson arrival of mRNA bursts. Thus non-zero values of $\mathcal{D}_{mp}$ indicate non-Poisson arrival of mRNA bursts.
Interestingly, for this condition there is 
no need to assume that the mRNA burst distribution is geometric. That is, the condition holds true for arbitrary 
mRNA burst distributions. Also, the condition does not require measurement of third moments.

\subsection*{Signatures for a simple kinetic scheme}
To illustrate the prescription derived for determining non-Poisson arrival processes, we consider a
specific kinetic scheme, Fig. 2a. 
For this kinetic scheme, the mRNA arrival time distribution in the Laplace domain is given by ( Eq. (\ref{efs}) in Appendix B) 
\begin{equation}{\label{efs2s}}
 f_L(s)=\frac{k_m \left(\alpha+s\right)}{k_m \left(\alpha+s\right)+s
   \left(\alpha+\beta+s\right)}.
\end{equation}
Using this in Eq.(\ref{ekg}) we find the gestation factor, $K_g$, and hence the mean, Fano factor and skewness
for both mRNAs and proteins as discussed in sec \ref{sec:mapping}. Finally, we derive exact analytic expressions for  $\mathcal{D}_m$, $ \mathcal{D}_p$ and $\mathcal{D}_{mp}$ 
from Eqs. (\ref{econditionM}), (\ref{econditionP}) and (\ref{econditionMP}) respectively. The expression for $\mathcal{D}_m$ reads
\begin{equation}
 \mathcal{D}_m=\frac{2k_m\beta(1-\langle m_b \rangle)\theta\left(1+\frac{(1+\alpha)(\alpha+\beta)\langle m_b \rangle k_m}{\theta(\langle m_b \rangle-1)}\right)}
                {(2+\alpha+\beta)(\theta+\beta k_m)((2\langle m_b\rangle-1)\theta+2k_m\langle m_b \rangle \beta)},
\end{equation}
where
\begin{equation}
 \theta=(\alpha+\beta)(\alpha+\beta+1),
\end{equation}

and we have set $\mu_m=1$ for simplicity. As expected, we note that $\mathcal{D}_{m}$ vanishes for the Poisson arrival processes,
i.e., either when $\beta$ is zero, or when the switching rates $\alpha$ and $\beta$ are very large compared to the rate of transcription, $k_m$.
The general expression for $\mathcal{D}_p$ is complicated. However, to gain insight about the arrival process, we can write down a simpler expression for $\mathcal{D}_p$ in the burst limit, $\mu_m=1\gg\mu_p$: 
\begin{equation}
 \mathcal{D}_p=-\frac{2\langle m_b\rangle^2 k_m^2k_p^2\alpha\beta}{(\alpha+\beta)^4+3\langle m_b\rangle k_p(\alpha+\beta)^2
               \psi+2\langle m_b\rangle^2 k_p^2\psi^2},
\end{equation}
where
\begin{equation}
 \psi=k_m\beta+(\alpha+\beta)^2.
\end{equation}
Again, for Poisson arrival processes $\mathcal{D}_p$ vanishes.
Finally, we obtain an analytic expression for  $\mathcal{D}_{mp}$, which is given by
\begin{eqnarray}
 \mathcal{D}_{mp}=\frac{\beta \left(\mu _p-\mu _m\right)}{\alpha \left(\alpha+\beta+\mu _m\right)
   \left(\alpha+\beta+\mu _p\right)},
\end{eqnarray}
and as expected, we note that $\mathcal{D}_{mp}$ vanishes for Poisson arrivals
and is negative for $\mu_p<\mu_m$. 
In Fig. 2c, we have plotted  all the three quantities together with simulation results as a function of $\beta$.

\section{Estimation of burst parameters}
\label{sec:burst}
The conditions derived above can be used to determine if the burst arrival process is non-Poisson. 
If this is indeed the case, then it is no longer accurate to estimate burst parameters based on 
measurements of mean and variance only, as has been done in previous studies \cite{weinberger2012expression}.
In the following, we present approaches for estimating burst parameters in this general case.

We start by considering the general kinetic scheme shown in Fig.\ref{fig:scheme}.  
This form for the kinetic scheme is supported by recent experiments in mammalian cells which suggest the presence 
of multiple rate-limiting steps between transition of the promoter from OFF to ON state \cite{suter2011mammalian,abhiBioinfoNew}. However, as observed in 
these experiments, a promoter in the ON state switches to the OFF state by a single rate-limiting step. We model the 
promoter switching from OFF to ON state by a general waiting-time distribution, $g(t)$. The switching rate from ON  
to OFF state is given by $\beta$. 
 \begin{figure}[h]
\centering
 \includegraphics[width=7cm]{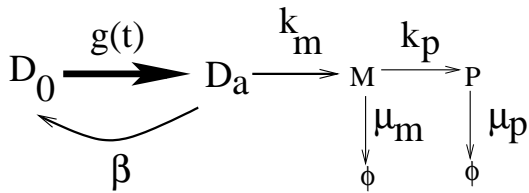}
\caption{Schematic representation of the general kinetic scheme. Thick line from inactive state $D_0$ to active state $D_a$
 represents a general kinetic scheme  with $g(t)$ as the waiting time distribution for the promoter to switch ON. }  
\label{fig:scheme}
\end{figure}

\subsection{Burst parameters from the sequence-size function}
To extract burst parameters for the general scheme considered above, we first note that bursts are 
generated due to the interplay of two time scales, one that corresponds to production of mRNAs (when the gene is active) 
while the other one corresponds to the waiting-time between production events (when the gene is in inactive state). For bursty gene expression, we expect a clear separation of time-scales between the characteristic time periods for these two cases. 
Following \cite{dobrzynski2009elongation}, it is convenient to define a sequence-size function, 
\begin{equation}{\label{ebp1}}
 \phi(\tau)=\frac{1}{1-\int_{0}^{\tau}f(t)dt},
\end{equation}
where $f(t)$ is the waiting-time distribution for the arrival of {\em single} mRNAs starting with the promoter in the ON state. For a fixed $\tau$,
the sequence-size function categorizes time intervals larger than $\tau$ as separating bursts. Correspondingly,
The term $1-\int_{0}^{\tau}f(t)dt$ represents the fraction of all mRNA arrivals that correspond to the arrivals produced 
in a single burst; thus $\phi$ provides the corresponding mean burst size.
For bursty gene expression with a separation of time-scales, for a specific choice of $\tau = \tau_x$, the 
sequence-size function can be related to the actual mean burst size. If $f(t)$ can be measured, then determination of $\tau_x$ can result in 
accurate estimates of the burst parameters such as mean burst size and frequency. In the following, we discuss how 
to determine $\tau_x$ for the general class of arrival processes considered in Fig. \ref{fig:scheme}.

The key insight is based on the observation that, due to the separation 
of time scales within bursts and between consecutive bursts, determination of $\tau_x$ 
can be done by using a simple two-state model as shown in Fig.2a. 
Even though the actual waiting time distribution between bursts ($g(t)$) may differ from the exponential distribution
for the two-state model, the short-time behavior of the sequence-size function will be indistinguishable 
between the two cases (given separation of time-scales). If $\tau_x$ can be connected to the short-time behavior, then analytical expressions for the sequence-size function
$\phi(\tau)$ for the two-state model can be used to estimate $\tau_x$ and thereby the mean burst size.
For the two-state model, we find that burst size can be determined using a specific $\tau_x$, 
which corresponds to an inflexion point where the curvature of $\phi(\tau)$ changes its sign. 
Specifically, for the two-state model, we obtain $f(t)$ by taking inverse Laplace transform of $f(s)$ given 
by Eq.(\ref{efs2s}). In the burst-limit, i.e.,  $\alpha/\beta\rightarrow 0$, we find its sequence function using Eq.~(\ref{ebp1}), 
and is given as
 \begin{eqnarray}
  \phi(\tau)=\frac{\left(k_m+\beta \right) e^{\tau \left(k_m+\beta \right)}}{k_m+\beta  e^{\tau \left(k_m+\beta \right)}},
 \end{eqnarray}
 and the value of $\tau$ at which $\phi(\tau)$ exhibits inflexion is
\begin{equation}
 \tau_x=\frac{1}{k_m+\beta}\ln \frac{k_m}{\beta}, \hspace{0.75cm} k_m>\beta.
\end{equation}
 The sequence size function $\phi(\tau)$ at this point ($\tau=\tau_x)$ is given by:
\begin{equation}
 \phi(\tau_x)=\frac{1}{2}\left(1+\frac{k_m}{\beta}\right)=\frac{1}{2}\left(1+\langle m_b \rangle\right),
\end{equation}

Thus, the procedure for determination of the mean burst size $(1+\langle m_b \rangle)$, given $f(t)$, is as follows:
\begin{enumerate}
\item Obtain the sequence-size function $\phi(\tau)$ from $f(t)$. For bursty synthesis, $\phi(\tau)$ will have 
an inflexion point.
 \item The mean burst size $(1+\langle m_b \rangle)$ is simply twice the value of the 
the sequence-size function at the inflexion point, $\tau_x$.
\end{enumerate}
This approach has been validated using 
stochastic simulations for multiple promoter models with correspondingly complex waiting-time distributions
between bursts (See Fig. 4).

\subsection{Estimation of $f(t)$ from steady-state moments}
The procedure outlined in the previous section assumes that the waiting-time distribution 
$f(t)$ can be obtained. However, this can be challenging experimentally, thus it is desirable 
to develop approaches for estimating $f(t)$ based on measurements of steady-state distributions.

To proceed in this direction, let us first obtain a relation connecting the two waiting-time distributions 
$f(t)$ (for single mRNA arrival) and $g(t)$ (for burst arrival). 
In Fig. \ref{fig:scheme}, we note that when the promoter is in the active
state, $D_a$, it can make multiple trips to $D_0$ before producing mRNA. Whenever gene is in $D_a$ state, it can 
either create mRNA or can switch back to $D_0$ state. The life-time of active state is a Poisson random variable
with distribution $(k_m+\beta)e^{-(k_m+\beta)t}$. Gene in $D_a$ state can produce mRNA either in a single step, i.e., without switching
back to $D_0$ state, or by making multiple trips to $D_0$ before producing mRNA. 
Denoting the number of trips made before producing mRNA by $q$, we obtain
that the Laplace transform of the waiting-time distribution $f(t)$ is given 
by 
\begin{equation}
 f_L(s)=\frac{k_m}{\beta+k_m}\sum_{q=0}^{\infty} \left(\frac{\beta}{\beta+k_m}\right)^q \left[g_L(s)\right]^{q}\left(\frac{k_m+\beta}{k_m+\beta+s}\right)^{q+1},
\end{equation}
which leads to:
\begin{equation}{\label{eq_fs}}
 f_L(s)=\frac{k_m}{k_m+s+\left[1-g_L(s)\right]\beta}.
\end{equation}

In order to determine $f_L(s)$, we will assume a specific functional form for $g_L(s)$.  
We consider that $g_L(s)$ is given by the following rational function,
\begin{eqnarray}{\label{egLs}}
 g_L(s)\equiv g_n^m(s)=\frac{1+a_1s+a_2s^2\dots a_ms^m}{1+b_1s+b_2s^2\dots b_ns^n},\hspace{ 0.2cm} n>m.
\end{eqnarray}
This form for the Laplace transform of the waiting-time distribution is consistent with known waiting-time distributions 
for phase-type processes \cite{Ross:2006:IPM:1197141}
and thus is valid quite generally.

Once we have an explicit form for $f_L(s)$, the next step is to determine the parameters, $k_m$, $\beta$, $a_1\dots a_m$, and $b_1\dots b_n$.
Thus, in general, we need $m+n+2$ measurements to estimate these parameters if we use $g(s)=g_n^m(s)$.
The simplest case, $g_L(s)=g_{1}^{0}(s)$, implies the presence of one kinetic step from inactive state to active state, with rate $1/b_1$,
and so it corresponds to the standard two-state random telegraph model. For this simple kinetic scheme, we can find the parameters, $k_m$, $\beta$, and $b_1$,
and hence $f_L(s)$ and the sequence size function by using three measurements associated with either mRNAs or proteins. In Fig. 4a, using the first three moments of mRNAs, we 
have shown the variation of estimated sequence size function with time for three different values of $\beta$. The fact that data is bursty is reflected by the presence of inflection 
point as shown in Fig.\ref{fig:mb-2s} where we have plotted the variation of second derivative of $\phi$ with time. 
\begin{figure}
\centering
\includegraphics[width=8.5cm]{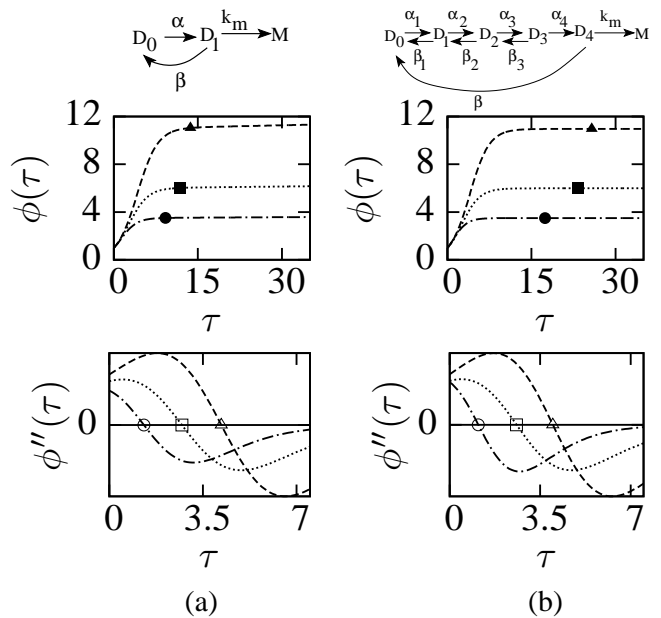}
\caption{Estimation of mean burst size from sequence size function $\phi(\tau)$ for two different transcriptional schemes, shown on the 
top panel of (a) and (b). The middle and bottom panels show the variations of $\phi(\tau)$ and $\phi^{\prime\prime}(\tau)$ as a function 
of time $\tau$ (scaled by $10^3$). The three lines correspond to three different values of $\beta$, 50 (dashed line), 100 (dotted line) and 
200 (dashed-dotted line), while keeping $k_m=500$. Other parameters: In (a) $\alpha=1$, while in (b),
$\alpha_1=1,\alpha_2=0.5,\alpha_3=0.25,\alpha_4=0.75,\beta_1=0.1,\beta_2=0.2,\beta_3=0.5$.
Estimated mean burst size has been shown by filled symbols while the observed inflexion 
points in the sequence size function are shown by empty symbols.}
\label{fig:mb-2s}  
\end{figure}

The form, $g_L(s)=g_{1}^{0}(s)$, is exact for the two-state random telegraph model. 
Using the expressions obtained for the first four steady-state moments, 
we can derive an analytic condition that determines whether the underlying mechanism can be represented by $g_{1}^{0}(s)$ (see Appendix).
However, if the arrival process is complex and involves multiple rate-limiting steps, then $ g_{1}^{0}(s)$ will not be an accurate
representation of the underlying kinetic process. In such cases, we need to use $g_L(s)$ of higher order.
The next step in this iterative process is to 
take $g_L(s)= g_{2}^{0}(s)$. This form of $g_L(s)$ is valid if there are only two rate-limiting steps in the promoter transition 
from OFF to ON state. For kinetic schemes that involve more than two steps, it will serve as an approximate reduced representation.
Interestingly, it turns out that even if $g_L(s)= g_{2}^{0}(s)$ is not the correct representation of the underlying kinetic 
process, this reduced representation works very well as far as estimating burst size is concerned. Alternatively, this
also means that the mean time for the arrival of mRNA bursts based on reduced representation matches with that of the underlying 
actual representation of the corresponding kinetic scheme. In Fig. 4b, we have illustrated the effectiveness of this approach for a
complex kinetic scheme for the promoter transition from OFF to ON state.

While the reduced representation, $g_L(s)=g^0_2(s)$, works reasonably well for estimating burst
size, with additional data, it is possible to extend the process further.
The iterative procedure we propose is as follows: 
\begin{enumerate}
\item Start with the simplest form $g_1^0(s)$ and
use three moments associated with either mRNA or proteins (or both) to find $f_L(s)$ as discussed above. Then this $f_L(s)$ can be
 used to get analytic predictions for higher moments \cite{liu00}.
\item If these analytic predictions
are consistent with the corresponding experimental observations then $g_1^0(s)$ provides a reasonable representation of the underlying kinetic scheme,
else a representation using more complex kinetic schemes is required. 
\item To address more complex kinetic schemes, we iteratively change $g_L(s)$ 
from $g_1^0(s)$ to $g_2^0(s)$, $g_3^0(s)$ \dots and so on, and iterate the steps outlined to determine the underlying $f_L(s)$.
However, we note that for uncovering more complex kinetic scheme we need additional measurements to estimate $f_L(s)$. 
If moment measurements are possible at {\emph different} mRNA/protein degradation rates, then these additional measurements can be used to estimate $f_L(s)$ and hence the corresponding mean transcriptional burst size. 
\end{enumerate}

\section{Discussion}
\label{sec:conclusions}
In this paper we study stochastic gene expression models with a general renewal-type arrival process for mRNAs. By mapping such a 
generic model of gene expression to systems studied in queueing theory, we derive analytical expressions for  the moments for mRNA and protein steady-state distributions. 
The expressions derived for  
these moments can be used to infer if the arrival process for mRNAs is non-Poisson. We have correspondingly obtained
analytic conditions that provide a signature for non-Poisson arrivals of mRNA. These conditions are in terms
of experimentally measurable quantities and can be tested by using measurements of either mRNA steady-state distributions or protein steady-state distributions or both, thus providing flexibility in terms of the availability of experimental data.
Apart from obtaining insights into the statistics of the arrival process, we can use the results derived for steady-state moments for accurately 
estimating burst parameters using an iterative approach.
It is hoped that future efforts based on these results, combining experiments and theoretical approaches, will be used in obtaining accurate representations of 
the arrival process and burst parameters for a wide range of cellular systems. 
 \begin{acknowledgments}
 This work was supported by the NSF through the grants PHY-1307067 and DMS- 1413111.
 \end{acknowledgments}
 
\appendix
\section{Derivation of steady-state moments for mRNAs and proteins}
\label{appendixA}
In this section, we discuss the derivation of expressions for the moments for mRNAs and protein steady-state distributions based 
on mapping of gene expression model to  $GI^X/M/\infty$ model in the queuing theory. For $GI^X/M/\infty$ model, with $\lambda$ and $\mu$ as the rates of customers mean arrival and service time respectively, 
exact steady state moments for the number of customers, $N$, can be obtained. Following\cite{liu00}, the 
binomial moments are given by 
\begin{equation}{\label{eAGk}}
 \left\langle \prod_{i=1}^k\left(X-i+1\right)\right\rangle=G^k(1),
\end{equation}
 where the symbol $\langle~\rangle$ stands for average over many ensembles, $G^k(1)$ corresponds to $k^{\text{th}}$ differentiation of $G(z)$ with respect to $z$ at $z=1$,
with $$G(z)=1+\sum_{r=1}^{\infty}B_r(z-1)^r.$$ The coefficients $B_r$ are given as:
\begin{equation}{\label{eABr}}
 B_r=\lambda\sum_{k=1}^{r}\frac{A_k}{k!}B^{\star}_{r-k}(k\mu),  
\end{equation}
where
\begin{equation}{\label{eABs}}
 B^\star_r(s)=\frac{f_L(s)}{1-f_L(s)}\sum_{k=1}^r\frac{A_k}{k!}B^\star_{r-k}(s+k\mu),
\end{equation}
and  $A_k=d^kA(z)/dz^k|_{z=1}$, with $A(z)$ as the generating function for 
the batch size distribution of arriving customers
and $f_L(s)$ is the Laplace transform of arrival time distribution of customers, $f(t)$.
Using $B_0^\star(s)=1/s$, we can iteratively find all the $B_r$ coefficients using 
 Eqs.~(\ref{eABr}),(\ref{eABs}) and hence all the moments from Eq.~(\ref{eAGk}).
Using this procedure, expressions for the moments of number of customers 
can be obtained explicitly.
For example, the corresponding
mean, $\langle N\rangle$, variance, $\sigma^2$, and skewness $\gamma$,
are given by:
\begin{eqnarray}{\label{eAmoments}}
 \langle N\rangle&=&\frac{\lambda}{\mu}A_{1},\nonumber\\
\frac{\sigma^2}{\langle N \rangle^2}&=&\frac{1}{\langle N \rangle}\left[1+A_1\left( \frac{K_g(\mu)-1}{2}+\frac{\lambda}{\mu}\right)
+\frac{A_2}{2A_1}-\langle N \rangle\right],\nonumber\\
\frac{\gamma\sigma^3}{\langle N \rangle}&=&1+2A_1^2\left(\frac{\lambda}{2\mu}\mathcal{K}_1(\mu)+\mathcal{K}_2(\mu,A_1)\right)
+A_2\mathcal{K}_3(\mu,A_1)\nonumber\\&&+\frac{A_3}{3A_1},
\end{eqnarray} 
where $\mathcal{K}_1,\mathcal{K}_2$, and $\mathcal{K}_3$ are three functions given by
\begin{eqnarray}{\label{eAkappa}}
 \mathcal{K}_1(x)&=&K_g(2x)-K_g(x),\nonumber\\ 
 \mathcal{K}_2(x,y)&=&\frac{K_g(x)-1}{4}\left(\frac{3}{y}+K_g(2x)-1\right),\nonumber\\
 \mathcal{K}_3(x,y)&=&\frac{3}{2y}+\frac{K_g(x)+K_g(2x)}{2}-1,
 \end{eqnarray}
and  $K_g(\mu)$ is the gestation factor, 
\begin{equation}{\label{eAkg}}
K_g(\mu)=1+2\left[\frac{f_L(\mu)}{1-f_L(\mu)}-\frac{\lambda}{\mu}\right],
\end{equation}
that encodes information about the arrival process. Extending this approach, we obtain expressions for higher moments. For example, fourth central moments are given by:
\begin{eqnarray}
 \langle(N-\langle N\rangle)^4\rangle&=24B_4+36B_3+14B_2+B_1+6B_1^2(B_1\left.\right.\nonumber\\&+2B_2)-4B_1\left(B_1+6(B_2+B_3)\right)-3B_1^4,\nonumber\\
\end{eqnarray}
where,
\begin{eqnarray}
 B_1&=&\frac{\lambda}{\mu}A_1,\nonumber\\
B_2&=&\frac{\lambda}{2\mu}\left(A_1^2\phi(\mu)+\frac{A_2}{2}\right),\nonumber\\
B_3&=&\frac{\lambda}{3\mu}\left[\frac{A_3}{6}+\frac{A_1A_2}{2}\left(\phi(\mu)+\phi(2\mu)\right)+A_1^3\phi(\mu)\phi(2\mu)\right],\nonumber\\
B_4&=&\frac{\lambda}{96\mu}\left[A_4+4A_1A_3\phi(3\mu)+6A_2\phi(2\mu)\left(A_2
       +2A_1^2\phi(3\mu)\right)\right.\nonumber\\&+&\left. 4A_1\phi(\mu)\left\{A_3+3A_1A_2\phi(3\mu)
+3A_1\phi(2\mu)\left(A_2+\right.\right.\right.\nonumber\\ && \left.\left.\left.2A_1^2\phi(3\mu)\right)\right\}\right]\nonumber\\
\end{eqnarray}
with $\phi(\mu)$ as
\begin{equation}
 \phi(\mu)=\frac{f_L(\mu)}{1-f_L(\mu)}.
\end{equation}

Eqs.(\ref{eAGk}),(\ref{eABr}) and (\ref{eABs}) can be used to derive steady state moments for mRNAs and proteins by mapping gene 
expression model to $GI^X/M/\infty$ model in queueing theory. 
To derive these expressions, we need to have the parameters associated  with  bursts statistics, $A_k^{m},A_k^{p}$, $k=1,2,3\dots$,
for both mRNAs and proteins, with superscripts '$m$' and '$p$' standing for mRNA and protein, respectively.
For mRNAs, we note that the burst size parameters $A_{1},A_{2}$ and $A_{3}$ are given by
\begin{eqnarray}{\label{eAms}}
A_1^m&=&\langle m_b\rangle,\nonumber\\
A_2^m&=&\langle m_b(m_b-1)\rangle,\nonumber\\
A_3^m&=&\langle m_b(m_b-1)(m_b-2)\rangle,
\end{eqnarray} 
where $m_b$ is the mRNA burst size. Using Eq. (\ref{eAms}) in Eqs. (\ref{eAmoments}) and (\ref{eAkappa}), we can write explicit expressions for 
the first three moments of mRNAs copy numbers as written in the main text, which are exact for all parameter ranges.

To obtain corresponding burst size parameters
for proteins ($A_k^p$), we note that each mRNA produces a random number of proteins, $p_b$. 
Using Eq. (1) in the main text, we can obtain expressions for the parameters $A_k^p$, which are given by
Eq. (\ref{ekp}).
Corresponding expressions for 
the first three moments of protein copy numbers has been written in the main text. It is to be 
noted that resulting expressions for protein variance and skewness are exact in the burst limit (i.e. $\mu_m\gg\mu_p$), however, 
beyond this limit one can write approximate expressions for these quantities, as illustrated in the main text.

It is also possible to obtain expressions for 
the fourth central moments of both mRNAs and proteins by using 
\begin{equation}
A_4^{m}=\langle m_b(m_b-1)(m_b-2)(m_b-3)\rangle
\end{equation}
 for     
mRNAs and, 
\begin{eqnarray}
A_4^{p}&=&3\left[\langle m_b(m_b-1)\rangle\langle p_b \rangle\langle p_b(p_b-1)(p_b-2)\rangle+\langle m_b(m_b-1)\rangle\right.\nonumber\\&&\left.
     \langle p_b(p_b-1)\rangle^2+\langle m_b(m_b-1)(m_b-2)\rangle \langle p_b \rangle^2 \langle p_b(p_b-1)\rangle\right]\nonumber\\&&
   +\langle m_b(m_b-1)(m_b-2)(m_b-3) \rangle \langle p_b\rangle^4 
\nonumber\\&&+3\langle m_b(m_b-1)(m_b-2)\rangle \langle p_b\rangle^2 \langle p_b(p_b-1)\rangle\nonumber\\&&+\langle m_b(m_b-1)\rangle
\langle p_b\rangle \langle p_b(p_b-1)(p_b-2)\rangle\nonumber\\&&+\langle m_b\rangle \langle p_b(p_b-1)(p_b-2)(p_b-3)\rangle,\nonumber\\
\end{eqnarray}
for proteins, by using Eq. (\ref{gen_fn_eqn}). Again, the fourth moment for mRNAs is exact for the entire parameter regime, while for proteins the expression
is exact only in 
the burst limit, i.e. $\mu_m\gg\mu_p$. Beyond the burst limit, we can find approximate expressions for the fourth moment using 
the approach outlined for the second and third moments, i.e. comparing our exact result in 
the burst-limit with the exact results obtained for the two-stage model\cite{bokes2012exact}. This leads to the corresponding scaling of the
coefficients $A_k^p$, specifically:
\begin{eqnarray}
  &&A_2^p\rightarrow A_2^p\frac{1}{1+\frac{\mu_p}{\mu_m}},\nonumber\\
  &&A_3^p\rightarrow A_3^p\frac{1}{(1+\frac{\mu_p}{\mu_m})(1+2\frac{\mu_p}{\mu_m})},\nonumber\\
  &&A_4^p\rightarrow A_4^p\frac{1}{(1+\frac{\mu_p}{\mu_m})(1+2\frac{\mu_p}{\mu_m})(1+3\frac{\mu_p}{\mu_m})}.
\end{eqnarray}
As shown in Fig.\ref{fig:4th} the resulting analytical expression shows good agreement with results from simulations.
\begin{figure}
\centering
 \includegraphics[width=5cm]{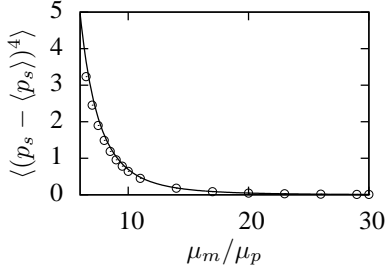}
\caption{Steady state fourth central moment of proteins (scaled by $10^{-11}$) for the model shown in Fig. 2a of the main text. Here lines represent analytic estimates
while points correspond to the simulation results. Parameters are: $\alpha=0.5$, $\beta=0.25$, $k_m=2$, $\langle m_b \rangle=5$, $k_p=0.5$. }
\label{fig:4th}
\end{figure}
\section{Illustrative examples for condition identifying non-geometric bursts}
In this section, we consider illustrative examples for the condition relating to 
the assumption of geometric burst distribution for mRNAs. 
\subsubsection{Poisson arrival of negative binomial bursts}
For Poisson arrival of negative binomial bursts, given by Eqs. (\ref{n1}) and (\ref{n2}) in the main text,
let us first consider the steady state expressions
for the moments. Using Eq. (\ref{n2})
we note that,
\begin{eqnarray}{\label{n3}}
 \langle m_b \rangle &=& \frac{p r}{1-p},\nonumber\\
 \langle m_b(m_b-1) \rangle  &=& \frac{p^2 r (r+1)}{(p-1)^2},\nonumber\\
\langle m_b(m_b-1)(m_b-2)\rangle &=& -\frac{p^3 r (r+1) (r+2)}{(p-1)^3}.
\end{eqnarray}
Plugging these values in Eqs.~(\ref{emeanMP}),(\ref{enoiseM}) and (\ref{eskewM}) of main text
and making use of Eq.~(\ref{n1}) for $f_L(s)$, 
we obtain the expression for the steady state moments. For example,  mean number of mRNAs can be written as
\begin{equation}{\label{n4}}
 \langle m_s \rangle=\frac{k_b p r}{\mu_m(1- p)},
\end{equation}
its Fano factor as
 \begin{equation}{\label{n5}}
  F_m=\frac{p (r-1)+2}{2(1-p)},
 \end{equation}
 and its skewness as
\begin{equation}{\label{n6}}
 \frac{\gamma_{m_s}\sigma^3_{m_s}}{\langle m_s \rangle}=\frac{p (p (r-1) (2 r-1)+9 r-3)+6}{6 (p-1)^2}.
\end{equation}
Using these moments in Eq.~(\ref{eGm}), we get an explicit expression for  $\mathcal{G}_m$:
\begin{equation}
 \mathcal{G}_m=\frac{1}{3} \left(-\frac{p+1}{p r+1}+\frac{4}{p (r-1)+2}+2\right),
\end{equation}
as written in the main text.
We notice that for the geometric bursts ($r=1$) we get $\mathcal{G}_m=1$, as expected. However, away from this limit ($r=1$), deviations
of $G_m$ values away from 1 can be seen, see Fig. \ref{fig:nb}.
\begin{figure}
\centering
 \includegraphics[width=5cm]{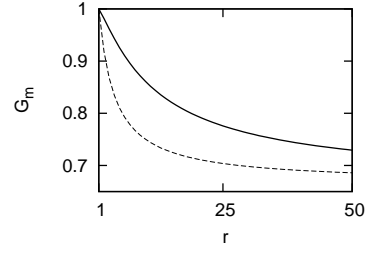}
\caption{$\mathcal{G}_m$ as a function of $r$ for two different values of $p$, 0.25 (solid line) and 0.75 (dashed line).}
\label{fig:nb}
\end{figure}

\begin{figure}[b]
 \centering
  \includegraphics[width=5cm]{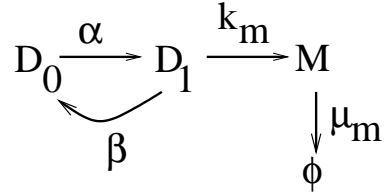}
 \caption{Schematic representation for the transcriptional kinetic scheme of two state model. Gene in 
OFF state ($D_0$) switches
to ON state ($D_1$) with rate $\alpha$ and can switch back to OFF state with rate $\beta$. When the gene is 
ON, it produces mRNA bursts with rate $k_m$, and mRNAs can then degrade with rate $\mu_m$.}
 \label{fig:2s}
 \end{figure}

\subsubsection{Two-state random telegraph model} Next, we consider the two-state random 
telegraph model, a widely used model for gene expression, Fig.~\ref{fig:2s}.
Here the gene switches stochastically between its ON and OFF states: the rate
of switching from ON to OFF is $\alpha$ while that from
OFF to ON it is $\beta$. Gene in the ON state
then produces a single mRNA with rate $k_m$, which can degrade further with
rate $\mu_m$. To verify our condition for geometric bursts, the first
step is to find mRNA moments, mean, Fano factor and skewness. However,
as can be seen in Eqs.~(4) and (7), to find
these moments the central quantity that needs to be evaluated is
$f_L(s)$, the waiting time distribution for the arrival of mRNA bursts
in the Laplace domain.  Equivalently, this waiting time distribution
translates into finding the first passage time distribution for the
production of mRNA given that gene is in the active state $D_1$ at
time $t=0$.  If $P_0(t)$ and $P_1(t)$ denote the probabilities
of gene being in OFF and ON states at time $t$,
respectively, then the first passage time distribution is given by,
\begin{equation}{\label{eft}}
 f(t)=k_mP_1(t),
\end{equation} 
where the probabilities, $P_0(t)$ and $P_1(t)$ obey the Master equation
\begin{eqnarray}{\label{epdot}}
 \frac{dP_0(t)}{dt}&=&\beta P_1(t)-\alpha P_0(t),\nonumber\\
\frac{dP_1(t)}{dt}&=&\alpha P_0(t)-\beta P_1(t).
\end{eqnarray}
The corresponding evolution equation in the Laplace domain is given by
\begin{eqnarray}{\label{efins}}
sf_0(s)-x_0&=&\beta f_1(s)-\alpha f_0(s),\nonumber\\
sf_1(s)-y_0&=& \alpha f_0(s)-\beta f_1(s),
\end{eqnarray}
where  $f_j(s)$  stands for the  Laplace transform of  $P_j(t)$, and $x_0$ and $y_0$ are the initial 
values of $P_0$ and $P_1$, respectively. For the process in the Fig. \ref{fig:2s}, where mRNAs are always produced 
from the active state, we take $P_0=0$ and $P_1=1$, and obtain the Laplace transform of first passage 
waiting time distributions as 
\begin{equation}{\label{efs}}
 f_L(s)=\frac{k_m(\alpha+s)}{s^2+s(\alpha+\beta+k_m)+\alpha k_m}.
\end{equation}
Using this $f_L(s)$ in Eqs.~(\ref{emeanMP}),(\ref{enoiseM}) and (\ref{eskewM}), we obtain explicit expressions for the
first three moments of mRNA copy numbers:  
\begin{eqnarray}{\label{eMmoments2s}}
 \langle m_s \rangle &=&\left(\frac{\alpha}{\alpha+\beta}\right)\frac{k_m}{\mu_m},\nonumber\\ 
  F_m&=&1+\frac{\beta k_m}{(\alpha+\beta)(\mu_m+\alpha+\beta)},\nonumber\\
  \frac{\gamma_{m_s}\sigma^3_{m_s}}{\langle m_s \rangle}&=&\frac{1}{(\alpha+\beta)^2(\mu_m+\alpha+\beta)(2\mu_m+\alpha+\beta)}
              \left[ \alpha^4+\right. \nonumber\\ &&\left. 4\alpha^3\beta +\beta^2(k_m+\beta)(2k_m+\beta)
  +2\mu_m^2(\alpha+\beta)^2 \right.\nonumber\\&& \left.
   +3\alpha^2\beta(k_m+2\beta)+2\alpha\beta(-{k_m}^2+3k_m\beta+
  2\beta^2)\right.\nonumber\\&& \left.+3\mu_m(\alpha+\beta)(2k_m\beta+(\alpha+\beta)^2)\right]
\end{eqnarray}
Using these values of mean, Fano factor and skewness in Eq.~(\ref{eGm}), we get $\mathcal{G}_m=1$, as expected.

\subsubsection{Transcription from two promoter states} Finally, we consider a model as shown in Fig \ref{fig:2on}. Here $D_0$,
$D_1$ and $D_2$ are three promoter states. Now, instead of having mRNA production from just a single state, as discussed above, let us 
assume that mRNAs are produced by two states $D_1$ and $D_2$ with rates $k_{m1}$ and $k_{m2}$, respectively.
In the absence of any one of these two transcriptional routes, bursts are geometrically produced as discussed
above. However, when both transcriptional routes are present we expect deviation from $\mathcal{G}_m=1$,
which we show in the following.  
  
To start with, let us first denote by $P_{\sigma}(m,t)$ as the probability that there  are $m$ number of mRNAs
at a time $t$ in the promoter state $\sigma=0,1,2$. The evolution of these probabilities reads as
\begin{eqnarray}{\label{e3p1}}
\frac{P_0(m,t)}{dt}&=&\mu_m(m+1)P_0(m+1,t)+\beta_1 P_2(m,t)\nonumber\\&-&\left(\alpha+\mu_m m\right)P_0(m,t),\nonumber\\
\frac{P_1(m,t)}{dt}&=&\alpha P_0(m,t)+k_{m1}P_1(m-1,t)+\mu_m(m+1)\nonumber\\&&P_1(m+1,t)-(\beta_2+k_{m1}+\mu_m m)P_1(m,t),\nonumber\\
\frac{P_2(m,t)}{dt}&=&\beta_2 P_1(m,t)+k_{m2}P_2(m-1,t)+\mu_m (m+1)\nonumber\\&&P_2(m+1,t)-\left(\beta_1+k_{m2}+\mu_m m\right)P_2(m,t).\nonumber\\
\end{eqnarray}
In the following, we will use this equation to get the first three moments of mRNA in the steady state. Let us first sum 
over all possible values of $m$ and use the  normalization $\sum_{\sigma}P_\sigma(m)=1$. This leads to  
\begin{eqnarray}
 P_0&=&\frac{\beta_1\beta_2}{\beta_1\beta_2+\alpha(\beta_1+\beta_2)},\nonumber\\
 P_1&=&\frac{\alpha\beta_1}{\beta_1\beta_2+\alpha(\beta_1+\beta_2)},\nonumber\\
 P_2&=& \frac{\alpha\beta_2}{\beta_1\beta_2+\alpha(\beta_1+\beta_2)}.
\end{eqnarray}
Next, multiplying Eq. (\ref{e3p1}) by $m$ and summing over all $m$, we have
\begin{eqnarray}
 &&\beta_1\langle m\rangle_2-(\mu_m+\alpha)\langle m\rangle_0=0,\nonumber\\
 &&\alpha\langle m \rangle_0+k_{m1}P_1-(\mu_m+\beta_2)\langle m\rangle_1=0,\nonumber\\
 &&\beta_2\langle m\rangle_1+k_{m2}P_2-(\mu_m+\beta_1)\langle m\rangle_2=0,
\end{eqnarray}
where $\langle m \rangle_{\sigma}=\sum_{m}mP_{\sigma}(m)$. These equations are solved to
get the mean number of mRNAs as  
\begin{equation}
 \langle m \rangle=\sum_{\sigma}\langle m \rangle_{\sigma}=\frac{k_{m1}P_1+k_{m2}P_2}{\mu_m}.
\end{equation}
Similarly, if we multiply Eq. (\ref{e3p1}) by $m^2$ and sum over all $m$, and denote $\langle m^2 \rangle_{\sigma}=\sum_m m^2P_{\sigma}(m)$, 
we get
\begin{eqnarray}
 &&\beta_1\langle m^2\rangle_2+\mu_m\langle m\rangle_0-(2\mu_m+\alpha)\langle m^2\rangle_0=0,\nonumber\\
&&\alpha\langle m^2\rangle_0-(\beta_2+2\mu_m)\langle m^2\rangle_1+(2k_{m1}+\mu_m)\langle m \rangle_1\nonumber\\&&+k_{m1}P_1=0,\nonumber\\
&&\beta_2\langle m^2\rangle_1-(\beta_1+2\mu_m)\langle m^2\rangle_2+(\mu_m+2k_{m2})\langle m_2\rangle\nonumber\\&&+k_{m2}P_2=0,\nonumber\\
\end{eqnarray}
which can be solved to get $\langle m^2\rangle=\sum_{\sigma}\langle m^2\rangle_{\sigma}$.
Finally, to get third moment we  multiply Eq. (\ref{e3p1}) by $m^3$ and sum over $m$, the 
resulting equations read
\begin{eqnarray*}
&&\mu_m\left[-3\langle m^3\rangle_0+3\langle m^2\rangle_0-\langle m\rangle_0\right]+\beta_1\langle m^3\rangle_2-\alpha\langle m^3\rangle_0=0,\nonumber\\
&&\alpha\langle m^3\rangle_0+k_{m1}\left[3\langle m^2\rangle_1+3\langle m\rangle_1+P_1\right]\nonumber\\&&+\mu_m\left[-3\langle m^3\rangle_1
+3\langle m^2\rangle_1-\langle m\rangle_1\right]
-\beta_2\langle m^3\rangle_1=0,\nonumber\\
&&\beta_2\langle m^3\rangle_1+k_{m2}\left[3\langle m^2\rangle_2+3\langle m\rangle_2+P_2\right]\nonumber\\&&+\mu_m\left[-3\langle m^3\rangle_2
+3\langle m^2\rangle_2-\langle m\rangle_2\right]
-\beta_1\langle m^3\rangle_2=0,\nonumber\\
\end{eqnarray*}
which are solved to get the third moment of mRNAs. Once we have the first three moments, we can evaluate $\mathcal{G}_m$ using Eq. (\ref{eGm}). The 
resulting expression is somewhat complicated, and therefore we just show the result in Fig. \ref{fig:2on}. As can be seen, for a given set of 
other parameters, variations of $\mathcal{G}_m$ with $k_{m2}$ show that it approaches 1 for $k_{m2}=0$, as expected. However, beyond this 
significant deviations are visible.   
\begin{figure}
 \includegraphics[width=5cm]{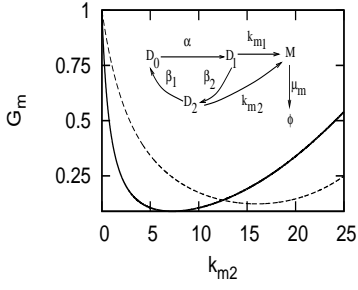}
\caption{Variation of $\mathcal{G}_m$ as a function of transcriptional rate 
$k_{m2}$ has been shown for two different values of $\alpha$, 1(solid line) and 2(dashed line), for the model (inset). Other parameters are: $\beta_1=0.5,\beta_2=0.25,k_{m1}=40,
\mu_m=1$.}
\label{fig:2on}
\end{figure}
\section{Condition for non-geometric bursts using protein steady-state moments}
The condition for geometric bursts derived in the paper is an exact
condition using mRNA steady state measurements, however, it is of
interest to see if a similar condition can be obtained using moments
of the protein steady-state distribution. As we show below, this
indeed can be done in the limit that the mRNA degradation rate is very
large compared to that of protein (which is a valid approximation in
many cellular systems). In this so called burst limit,  
we can derive analytical expressions for the first three moments of the protein steady-state 
distribution in the burst limit as discussed in the main text (Eqs. (\ref{emeanMP}), (\ref{enoiseP}), and (\ref{eskew1P})). 
In the burst limit, for geometrically distributed
 mRNA and protein bursts, the variance in protein copy numbers, 
$\sigma^2_{p_s}=\langle p_s^2 \rangle -\langle p_s \rangle^2$, is written as
\begin{equation}{\label{eAnoiseP}}
\frac{\sigma^2_{p_s}}{\langle p_s \rangle^2}=\frac{1}{\langle p_s \rangle}\left[1+\frac{b_{}}{2}\left(1+K_g(\mu_p)\right)\right],
\end{equation}
and similarly, skewness is given by
\begin{eqnarray}{\label{eskewP}}
 \frac{\gamma_{p_s}\sigma^3_{p_s}}{\langle p_s \rangle}&=& 1+2b_{}^2\left[\frac{\langle p_s\rangle}{2b_{}}\mathcal{K}_1(\mu_p)+\mathcal{K}_2(\mu_p,b_{})
\vphantom{\frac{1}{1+\frac{2}{3}}}+\mathcal{K}_3(\mu_p,b_{})+1\right].\nonumber\\
\end{eqnarray}
Using Eq. (\ref{eAnoiseP}) in (\ref{eskewP})
leads to
\begin{eqnarray}{\label{eGp}}
 \mathcal{G}_p&\equiv&\frac{\gamma_{p_s}\sigma^3_{p_s}\langle p_s \rangle^{-1}}{1
  +2\langle p_s\rangle\left(F_p(2\mu_p)-F_p\right)+(F_p-1)(1+2F_p(2\mu_p))}\nonumber\\&=&1
\end{eqnarray}
This is the condition for the protein burst distribution to be
geometric. Note that if the protein burst distribution is geometric,
this implies that the underlying mRNA burst distribution is a
conditional geometric distribution \cite{elgart2011connecting}. Thus the value of
$\mathcal{G}_p \neq 1 $ indicates that both protein and mRNA burst
distributions differ from the geometric distribution.

Next, we verify this condition for the two-state random telegraphic model of Fig.~\ref{fig:23moments}a,
with each transcription event leading to the arrival of just one mRNA instead of arrival of conditionally geometric 
bursts. 
For this, we first note that, using Eqs.~(2),(6), (9) and (\ref{efs}), protein moments 
are given by
\begin{eqnarray}{\label{ePmoments2s}}
 \langle p_s \rangle&=&\left(\frac{\alpha}{\alpha+\beta}\right)\frac{k_m k_p}{\mu_m\mu_p},\nonumber\\
 F_p&=&1+\frac{k_p}{\mu_m}\left[1+\frac{k_m\beta}{(\alpha+\beta)(\mu_p+\alpha+\beta)}\right],\nonumber\\
\frac{\gamma_{p_s}\sigma^3_{p_s}}{\langle p_s \rangle}&=&\frac{1}{\xi(\mu_p+\xi)(\mu_p-\xi)^2\mu_m^2}\left[(k_p+\mu_m)(2k_p+\mu_m)\right.\nonumber\\&&\left.
\xi(\mu_p+\xi)(\mu_p-\xi)^2 -k_mk_p(\mu_p-\xi)(3\mu_m(\mu_p+\xi)\right.\nonumber\\ &&+\left. 2k_p(\mu_p+2\xi)-2k_mk_p)\beta
+4k_m^2k_p^2\beta^2\right]
\end{eqnarray}
where $$ \xi=\alpha+\beta+\mu_p. $$
Using this in Eq.~(\ref{eGp}), we get $\mathcal{G}_p=1$ which is consistent with the fact that the protein bursts distribution is geometric. 

\section{Condition for the two-state random telegraph model}
In this section, we derive analytic conditions for validating the proposition that the underlying kinetic scheme can be 
represented by a two-state
random telegraph model.
For this model, the form $g_L(s)=g_{1}^{0}(s)=1/(1+b_1s)$ in the main text with $g_L(s)$ as the phase-type process, given by 
Eq.(\ref{egLs}), is exact.
Using the 
first three mRNA moments, we can estimate the parameters $b_1$, $\beta$, and $k_m$ in terms these moments,
and plugging these into the equation for the fourth moment, we derive:
\begin{figure}
\centering
 \includegraphics[bb=131 401 245 481,width=6cm]{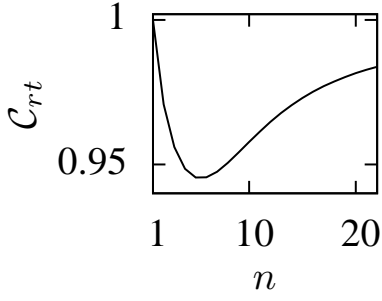}
\caption{Verification of two-state random telegraph model as the underlying kinetic scheme: $C_{rt}$ as a function of number of 
steps, $n$, for transitions from \emph{off} to \emph{on} state. Parameters: $\alpha=\beta=2$, $k_m=100$, $\mu=1$.}
\label{fig:Con2S}
\end{figure}
\begin{eqnarray}{\label{eCrt}}
 \mathcal{C}_{rt}&\equiv&  \frac{\nu_4^m\left(F_m (4 F_m+2 \langle m_s \rangle-5)-2 \langle m_s\rangle
-\frac{\nu_3^m}{\langle m_s \rangle}+2\right)}{\langle m_s\rangle\left(\xi_0+\xi_1 F_m+\xi_2 F_m^2+\xi_3 F_m^3\right)}\nonumber\\&=&1.
\end{eqnarray}
where $\nu_3^m=\langle(m_s-\langle m_s\rangle)^3\rangle$ and $\nu_4^m=\langle(m_s-\langle m_s\rangle)^4\rangle$ are the third and 
fourth central moments associated with mRNA measurements respectively, and $\xi_0, \xi_1, \xi_2 ~ \text{and}~  \xi_3$ are functions of 
$\langle m_s \rangle$ and $\nu_3^m$: 
\begin{eqnarray}
\xi_0&=&3\left(\frac{\nu_3^m}{\langle m_s\rangle}\right)^2(\langle m_s\rangle-3)+6\frac{\nu_3^m}{\langle m_s\rangle}(\langle m_s \rangle -1),\nonumber\\
\xi_1&=&3\left(\frac{\nu_3^m}{\langle m_s\rangle}\right)^2+(11-18\langle m_s\rangle)\frac{\nu_3^m}{\langle m_s\rangle}+4(\langle m_s\rangle-1),\nonumber\\
\xi_2&=&3\left(\frac{\nu_3^m}{\langle m_s\rangle}\right)(2+\langle m_s \rangle)-(6\langle m_s\rangle+13)\langle m_s\rangle+16,\nonumber\\
\xi_3&=& 6\langle m_s\rangle^2+15\langle m_s\rangle-17.
\end{eqnarray}
Thus once we have measurements of the first four moments associated with mRNA, then Eq.(\ref{eCrt}) must be satisfied if the 
underlying kinetic scheme is a two-state random telegraph model. That is,  Eq.(\ref{eCrt}) provides a prescription for the validity of 
two-state random telegraph model that can be tested experimentally.

For the sake of illustration, let us consider a kinetic scheme with $g_L(s)=(\alpha/(\alpha+s))^n$, which 
represents the presence of $n$ identical kinetic steps, each with rate $\alpha$, during the transition
of gene from OFF to ON. Using Eqs. (\ref{emeanMP}),(\ref{enoiseP}),(\ref{eskew1P}) and (\ref{efs}) we can explicitly find 
its moments that can be used in Eq. (\ref{eCrt}) to get $C_{rt}$. As expected, we can see that, in Fig. \ref{fig:Con2S}, 
for $n=1$, we get $C_{rt}=1$ and deviation from this value is evident for $n>1$.

\bibliographystyle{unsrt}
\bibliography{multi-WithGeo-prx-v5}

\begin{thebibliography}{10}

\bibitem{elowitz2002stochastic}
Michael~B Elowitz, Arnold~J Levine, Eric~D Siggia, and Peter~S Swain.
\newblock Stochastic gene expression in a single cell.
\newblock {\em Science}, 297(5584):1183--1186, 2002.

\bibitem{kaern2005stochasticity}
Mads K{\ae}rn, Timothy~C Elston, William~J Blake, and James~J Collins.
\newblock Stochasticity in gene expression: from theories to phenotypes.
\newblock {\em Nature Reviews Genetics}, 6(6):451--464, 2005.

\bibitem{raser2005noise}
Jonathan~M Raser and Erin~K O'Shea.
\newblock Noise in gene expression: origins, consequences, and control.
\newblock {\em Science}, 309(5743):2010--2013, 2005.

\bibitem{sanchez2013regulation}
Alvaro Sanchez, Sandeep Choubey, and Jane Kondev.
\newblock Regulation of noise in gene expression.
\newblock {\em Annual review of biophysics}, 42:469--491, 2013.

\bibitem{eldar2010functional}
Avigdor Eldar and Michael~B Elowitz.
\newblock Functional roles for noise in genetic circuits.
\newblock {\em Nature}, 467(7312):167--173, 2010.

\bibitem{raj2008nature}
Arjun Raj and Alexander van Oudenaarden.
\newblock Nature, nurture, or chance: stochastic gene expression and its
  consequences.
\newblock {\em Cell}, 135(2):216--226, 2008.

\bibitem{larson2011expression}
Daniel~R Larson.
\newblock What do expression dynamics tell us about the mechanism of
  transcription?
\newblock {\em Current opinion in genetics \& development}, 21(5):591--599,
  2011.

\bibitem{junker2014every}
Jan~Philipp Junker and Alexander van Oudenaarden.
\newblock Every cell is special: Genome-wide studies add a new dimension to
  single-cell biology.
\newblock {\em Cell}, 157(1):8--11, 2014.

\bibitem{munsky2012using}
Brian Munsky, Gregor Neuert, and Alexander van Oudenaarden.
\newblock Using gene expression noise to understand gene regulation.
\newblock {\em Science}, 336(6078):183--187, 2012.

\bibitem{golding2011decision}
Ido Golding.
\newblock Decision making in living cells: lessons from a simple system.
\newblock {\em Annual review of biophysics}, 40:63--80, 2011.

\bibitem{bar2006noise}
Arren Bar-Even, Johan Paulsson, Narendra Maheshri, Miri Carmi, Erin O'Shea,
  Yitzhak Pilpel, and Naama Barkai.
\newblock Noise in protein expression scales with natural protein abundance.
\newblock {\em Nature genetics}, 38(6):636--643, 2006.

\bibitem{newman2006single}
John~RS Newman, Sina Ghaemmaghami, Jan Ihmels, David~K Breslow, Matthew Noble,
  Joseph~L DeRisi, and Jonathan~S Weissman.
\newblock Single-cell proteomic analysis of s. cerevisiae reveals the
  architecture of biological noise.
\newblock {\em Nature}, 441(7095):840--846, 2006.

\bibitem{weinberger2012expression}
Leehee Weinberger, Yoav Voichek, Itay Tirosh, Gil Hornung, Ido Amit, and Naama
  Barkai.
\newblock Expression noise and acetylation profiles distinguish hdac functions.
\newblock {\em Molecular cell}, 47(2):193--202, 2012.

\bibitem{kumar2014exact}
Niraj Kumar, Thierry Platini, and Rahul~V Kulkarni.
\newblock Exact distributions for stochastic gene expression models with
  bursting and feedback.
\newblock {\em arXiv preprint arXiv:1409.3499}, 2014.

\bibitem{PhysRevX.4.041017}
Michael Hinczewski and D.~Thirumalai.
\newblock Cellular signaling networks function as generalized wiener-kolmogorov
  filters to suppress noise.
\newblock {\em Phys. Rev. X}, 4:041017, Oct 2014.

\bibitem{Balazsi2011Cell}
G{\'a}bor Bal{\'a}zsi, Alexander van Oudenaarden, and James~J Collins.
\newblock Cellular decision making and biological noise: from microbes to
  mammals.
\newblock {\em Cell}, 144(6):910--925, 2011.

\bibitem{suter2011origins}
David~M Suter, Nacho Molina, Felix Naef, and Ueli Schibler.
\newblock Origins and consequences of transcriptional discontinuity.
\newblock {\em Current opinion in cell biology}, 23(6):657--662, 2011.

\bibitem{coulon2013eukaryotic}
Antoine Coulon, Carson~C Chow, Robert~H Singer, and Daniel~R Larson.
\newblock Eukaryotic transcriptional dynamics: from single molecules to cell
  populations.
\newblock {\em Nature Reviews Genetics}, 14(8):572--584, 2013.

\bibitem{Golding20051025}
Ido Golding, Johan Paulsson, Scott~M. Zawilski, and Edward~C. Cox.
\newblock Real-time kinetics of gene activity in individual bacteria.
\newblock {\em Cell}, 123(6):1025 -- 1036, 2005.

\bibitem{chubb2006transcriptional}
Jonathan~R Chubb, Tatjana Trcek, Shailesh~M Shenoy, and Robert~H Singer.
\newblock Transcriptional pulsing of a developmental gene.
\newblock {\em Current biology}, 16(10):1018--1025, 2006.

\bibitem{raj2006stochastic}
Arjun Raj, Charles~S Peskin, Daniel Tranchina, Diana~Y Vargas, and Sanjay
  Tyagi.
\newblock Stochastic mrna synthesis in mammalian cells.
\newblock {\em PLoS biology}, 4(10):e309, 2006.

\bibitem{so2011general}
Lok-hang So, Anandamohan Ghosh, Chenghang Zong, Leonardo~A Sep{\'u}lveda, Ronen
  Segev, and Ido Golding.
\newblock General properties of transcriptional time series in escherichia
  coli.
\newblock {\em Nature genetics}, 43(6):554--560, 2011.

\bibitem{taniguchi2010quantifying}
Yuichi Taniguchi, Paul~J Choi, Gene-Wei Li, Huiyi Chen, Mohan Babu, Jeremy
  Hearn, Andrew Emili, and X~Sunney Xie.
\newblock Quantifying e. coli proteome and transcriptome with single-molecule
  sensitivity in single cells.
\newblock {\em Science}, 329(5991):533--538, 2010.

\bibitem{zong2010lysogen}
Chenghang Zong, Lok-hang So, Leonardo~A Sep{\'u}lveda, Samuel~O Skinner, and
  Ido Golding.
\newblock Lysogen stability is determined by the frequency of activity bursts
  from the fate-determining gene.
\newblock {\em Molecular systems biology}, 6(1), 2010.

\bibitem{sanchez2013genetic}
Alvaro Sanchez and Ido Golding.
\newblock Genetic determinants and cellular constraints in noisy gene
  expression.
\newblock {\em Science}, 342(6163):1188--1193, 2013.

\bibitem{dar2012transcriptional}
Roy~D Dar, Brandon~S Razooky, Abhyudai Singh, Thomas~V Trimeloni, James~M
  McCollum, Chris~D Cox, Michael~L Simpson, and Leor~S Weinberger.
\newblock Transcriptional burst frequency and burst size are equally modulated
  across the human genome.
\newblock {\em Proceedings of the National Academy of Sciences},
  109(43):17454--17459, 2012.

\bibitem{singh2012dynamics}
Abhyudai Singh, Brandon~S Razooky, Roy~D Dar, and Leor~S Weinberger.
\newblock Dynamics of protein noise can distinguish between alternate sources
  of gene-expression variability.
\newblock {\em Molecular systems biology}, 8(1), 2012.

\bibitem{gefen2008single}
Orit Gefen, Chana Gabay, Michael Mumcuoglu, Giora Engel, and Nathalie~Q
  Balaban.
\newblock Single-cell protein induction dynamics reveals a period of
  vulnerability to antibiotics in persister bacteria.
\newblock {\em Proceedings of the National Academy of Sciences},
  105(16):6145--6149, 2008.

\bibitem{weinberger2005stochastic}
Leor~S Weinberger, John~C Burnett, Jared~E Toettcher, Adam~P Arkin, and David~V
  Schaffer.
\newblock Stochastic gene expression in a lentiviral positive-feedback loop:
  Hiv-1 tat fluctuations drive phenotypic diversity.
\newblock {\em Cell}, 122(2):169--182, 2005.

\bibitem{Zeng2010682}
Lanying Zeng, Samuel~O. Skinner, Chenghang Zong, Jean Sippy, Michael Feiss, and
  Ido Golding.
\newblock Decision making at a subcellular level determines the outcome of
  bacteriophage infection.
\newblock {\em Cell}, 141(4):682 -- 691, 2010.

\bibitem{wernet2006stochastic}
Mathias~F Wernet, Esteban~O Mazzoni, Arzu {\c{C}}elik, Dianne~M Duncan, Ian
  Duncan, and Claude Desplan.
\newblock Stochastic spineless expression creates the retinal mosaic for colour
  vision.
\newblock {\em Nature}, 440(7081):174--180, 2006.

\bibitem{ochiai2014stochastic}
Hiroshi Ochiai, Takeshi Sugawara, Tetsushi Sakuma, and Takashi Yamamoto.
\newblock Stochastic promoter activation affects nanog expression variability
  in mouse embryonic stem cells.
\newblock {\em Scientific reports}, 4, 2014.

\bibitem{senecal2014transcription}
Adrien Senecal, Brian Munsky, Florence Proux, Nathalie Ly, Floriane~E Braye,
  Christophe Zimmer, Florian Mueller, and Xavier Darzacq.
\newblock Transcription factors modulate c-fos transcriptional bursts.
\newblock {\em Cell reports}, 8(1):75--83, 2014.

\bibitem{cai06}
L.~Cai, N.~Friedman, and X.~S. Xie.
\newblock Stochastic protein expression in individual cells at the single
  molecule level.
\newblock {\em Nature}, 440(7082):358--62, 2006.

\bibitem{Xie2006}
J.~Yu, J.~Xiao, X.~Ren, K.~Lao, and X.S. Xie.
\newblock Probing gene expression in live cells, one protein molecule at a
  time.
\newblock {\em Science}, 311(5767):1600--1603, 2006.

\bibitem{Pedraza2008}
J.M. Pedraza and J.~Paulsson.
\newblock Effects of molecular memory and bursting on fluctuations in gene
  expression.
\newblock {\em Science}, 319(5861):339--343, 2008.

\bibitem{zhang2014promoter}
Jiajun Zhang and Tianshou Zhou.
\newblock Promoter-mediated transcriptional dynamics.
\newblock {\em Biophysical journal}, 106(2):479--488, 2014.

\bibitem{peccoud1995markovian}
Jean Peccoud and Bernard Ycart.
\newblock Markovian modeling of gene-product synthesis.
\newblock {\em Theoretical population biology}, 48(2):222--234, 1995.

\bibitem{shahrezaei2008analytical}
Vahid Shahrezaei and Peter~S Swain.
\newblock Analytical distributions for stochastic gene expression.
\newblock {\em Proceedings of the National Academy of Sciences},
  105(45):17256--17261, 2008.

\bibitem{dobrzynski2009elongation}
Maciej Dobrzy{\'n}ski and Frank~J Bruggeman.
\newblock Elongation dynamics shape bursty transcription and translation.
\newblock {\em Proceedings of the National Academy of Sciences},
  106(8):2583--2588, 2009.

\bibitem{skupsky2010hiv}
Ron Skupsky, John~C Burnett, Jonathan~E Foley, David~V Schaffer, and Adam~P
  Arkin.
\newblock Hiv promoter integration site primarily modulates transcriptional
  burst size rather than frequency.
\newblock {\em PLoS computational biology}, 6(9):e1000952, 2010.

\bibitem{PhysRevLett.106.058102}
Tao Jia and Rahul~V. Kulkarni.
\newblock Intrinsic noise in stochastic models of gene expression with
  molecular memory and bursting.
\newblock {\em Phys. Rev. Lett.}, 106:058102, Feb 2011.

\bibitem{xu2013stochastic}
Xiaohua Xu, Niraj Kumar, Arjun Krishnan, and Rahul~V Kulkarni.
\newblock Stochastic modeling of dwell-time distributions during
  transcriptional pausing and initiation.
\newblock In {\em Decision and Control (CDC), 2013 IEEE 52nd Annual Conference
  on}, pages 4068--4073. IEEE, 2013.

\bibitem{suter2011mammalian}
David~M Suter, Nacho Molina, David Gatfield, Kim Schneider, Ueli Schibler, and
  Felix Naef.
\newblock Mammalian genes are transcribed with widely different bursting
  kinetics.
\newblock {\em Science}, 332(6028):472--474, 2011.

\bibitem{harper2011dynamic}
Claire~V Harper, B{\"a}rbel Finkenst{\"a}dt, Dan~J Woodcock, S{\"o}nke
  Friedrichsen, Sabrina Semprini, Louise Ashall, David~G Spiller, John~J
  Mullins, David~A Rand, Julian~RE Davis, et~al.
\newblock Dynamic analysis of stochastic transcription cycles.
\newblock {\em PLoS biology}, 9(4):e1000607, 2011.

\bibitem{KulkarniPRE2010}
V.~Elgart, T.~Jia, and R.V. Kulkarni.
\newblock Applications of littleÕs law to stochastic models of gene
  expression.
\newblock {\em Physical Review E}, 82(2):021901, 2010.

\bibitem{liu00}
L~Liu, B.~R.~K. Kashyap, and J.~G.~C. Templeton.
\newblock {On the GIX/G/Infinity system}.
\newblock {\em {Jour. Appl. Prob.}}, {27}({3}):{671--683}, {1990}.

\bibitem{bokes2012multiscale}
Pavol Bokes, John~R King, Andrew~TA Wood, and Matthew Loose.
\newblock Multiscale stochastic modelling of gene expression.
\newblock {\em Journal of mathematical biology}, 65(3):493--520, 2012.

\bibitem{ingram08}
Piers~J. Ingram, Michael P.~H. Stumpf, and Jaroslav Stark.
\newblock {Nonidentifiability of the Source of Intrinsic Noise in Gene
  Expression from Single-Burst Data}.
\newblock {\em {PLoS Comp Biol}}, {4}({10}), {2008}.

\bibitem{cookson2011queueing}
Natalie~A Cookson, William~H Mather, Tal Danino, Octavio
  Mondrag{\'o}n-Palomino, Ruth~J Williams, Lev~S Tsimring, and Jeff Hasty.
\newblock Queueing up for enzymatic processing: correlated signaling through
  coupled degradation.
\newblock {\em Molecular systems biology}, 7(1), 2011.

\bibitem{mather2010correlation}
William~H Mather, Natalie~A Cookson, Jeff Hasty, Lev~S Tsimring, and Ruth~J
  Williams.
\newblock Correlation resonance generated by coupled enzymatic processing.
\newblock {\em Biophysical journal}, 99(10):3172--3181, 2010.

\bibitem{little1961proof}
John~DC Little.
\newblock A proof for the queuing formula: L= $\lambda$ w.
\newblock {\em Operations research}, 9(3):383--387, 1961.

\bibitem{Ross:2006:IPM:1197141}
Sheldon~M. Ross.
\newblock {\em Introduction to Probability Models, Ninth Edition}.
\newblock Academic Press, Inc., Orlando, FL, USA, 2006.

\bibitem{bokes2012exact}
Pavol Bokes, John~R King, Andrew~TA Wood, and Matthew Loose.
\newblock Exact and approximate distributions of protein and mrna levels in the
  low-copy regime of gene expression.
\newblock {\em Journal of mathematical biology}, 64(5):829--854, 2012.

\bibitem{gillespie1977exact}
Daniel~T Gillespie.
\newblock Exact stochastic simulation of coupled chemical reactions.
\newblock {\em The journal of physical chemistry}, 81(25):2340--2361, 1977.

\bibitem{abhiBioinfoNew}
Petzold~L. Daigle~B., Soltani~M. and Singh A.
\newblock Inferring single-cell gene expression mechanisms using stochastic
  simulation.
\newblock {\em Submitted to Bioinformatics}, 2014.

\bibitem{elgart2011connecting}
Vlad Elgart, Tao Jia, Andrew~T Fenley, and Rahul Kulkarni.
\newblock Connecting protein and mrna burst distributions for stochastic models
  of gene expression.
\newblock {\em Physical biology}, 8(4):046001, 2011.

\end{thebibliography}
\end{document}